\documentclass[review,fleqn,sort&compress]{elsarticle}
\usepackage{amssymb,amsbsy,amsthm,amsmath,amsfonts,amssymb,amscd}
\usepackage{geometry}
\usepackage{babel,blindtext}
\usepackage{bm} 

\usepackage{graphicx}
\usepackage{float}
\usepackage{multirow}
\usepackage{subfigure}
\usepackage{lineno}
\usepackage{hyperref}
\hypersetup{colorlinks, citecolor=black, filecolor=black, linkcolor=black, urlcolor=black}
\usepackage{booktabs}
\usepackage[ruled,linesnumbered]{algorithm2e}
\usepackage{appendix}
\usepackage{algpseudocode}
\usepackage{mathtools,nccmath}
\usepackage{adjustbox}
\usepackage{cleveref}
\usepackage{dutchcal}
\usepackage[utf8]{inputenc}
\usepackage[scr=rsfso,cal=euler,frak=euler,bb=ams]{mathalfa}
\usepackage{longtable}
\usepackage{comment}
\usepackage{color}
\usepackage[dvipsnames]{xcolor}
\usepackage{cancel}
\usepackage{soul}
\colorlet{soulred}{red!40}

\usepackage{xparse}
\usepackage{mathrsfs}
\usepackage{upgreek}
\usepackage{colortbl}
\usepackage{hhline}
\newcolumntype{C}[1]{>{\centering\let\newline\\\arraybackslash\hspace{0pt}}m{#1}}

\begin{document}

\begin{frontmatter}
\title{Integrated Finite Element Neural Network (IFENN) for Phase-Field Fracture with Minimal Input and Generalized Geometry-Load Handling}

\author[NYUAD]{Panos Pantidis\corref{cor1}}
\author[UVM]{Lampros Svolos}
\author[NYUAD]{Diab Abueidda}
\author[NYUAD]{Mostafa E. Mobasher\corref{cor2}}
\cortext[cor1]{Corresponding author. \emph{E-mail address:} \texttt{pp2624@nyu.edu} (Panos Pantidis)}
\cortext[cor2]{Corresponding author. \emph{E-mail address:} \texttt{mostafa.mobasher@nyu.edu} (Mostafa Mobasher)}
\address[NYUAD]{Civil and Urban Engineering Department, New York University Abu Dhabi, Abu Dhabi, P.O. Box 129188, UAE}

\address[UVM]{Department of Civil and Environmental Engineering, University of Vermont, Burlington, VT, USA}

\begin{highlights}

    \item We propose a novel IFENN setup to model phase-field fracture propagation 

    \item IFENN is deployed with a physics-informed CNN (PICNN) without any temporal features

    \item PICNN trains only for 5 mins on strain energy–phase-field maps from just 2 load steps
    
    \item A single-crack-trained PICNN predicts the creation and nucleation of many cracks 

    \item Same PICNN is used across unseen geometries, loads, BCs and time-stepping schemes

\end{highlights}








\begin{abstract}

We present a novel formulation for modeling phase-field fracture propagation based on the Integrated Finite Element Neural Network (IFENN) framework. IFENN is a hybrid solver scheme that utilizes neural networks as PDE solvers within FEM, preserving accuracy via residual minimization while achieving speed-up via swift network predictions and reduction of the size of system of equations in coupled problems. In this work, we introduce a radically new formulation of IFENN in which the phase-field variable is calculated using physics-informed convolutional networks (PICNNs), while the equilibrium equation is still solved using FEM to maintain the solver robustness. Unlike conventional approaches, which rely on sequence or time-dependent models, we eliminate the need to include temporal features in the training setup and inference stage. Instead, we show that it is sufficient to learn only the spatial coupling between the strain energy density and the phase-field variable in the vicinity of the fracture process zone, and utilize this information along the advancing crack simulation. We train a single CNN in a purely physics-based, unsupervised manner on just two load increments from a single-notch tension problem, with a total training time of only 5 minutes. Following this exceptionally minimal and fast training, we show that the same PICNN can (when embedded within IFENN) model crack propagation in a very wide range of unseen scenarios, including arbitrarily rectangular domains, single and multiple interacting cracks, varying mesh densities, and arbitrary loading paths. The proposed formulation delivers breakthroughs that address many of the limitations in the existing literature of hybrid modeling, introducing a new paradigm for the development of generalizable, physics-consistent hybrid models that are applicable to fracture and other coupled problems.

\end{abstract}

\begin{keyword}
\texttt IFENN \sep phase-field fracture \sep convolution \sep PICNN \sep physics-informed \sep generalization 
\end{keyword}

\end{frontmatter}


\newpage
\section{Introduction}
\label{Section:Introduction}

\subsection{Literature review}

Most real-world mechanics-based problems are intrinsically multi-physics, typically encompassing non-linear interactions across multiple physical fields. Coupled phenomena often involve a broad range of interacting physical processes, each of which introduces its own set of complexities. From a computational standpoint, the solution of multi-physics problems still remains a notoriously challenging task, whether this is performed within the Finite Element Method (FEM) or other numerical approaches \cite{zhang2023thermal, li2024state, sedmak2018computational, rabczuk2013computational}. Typically, the associated challenges are tied to both the difficulty of achieving a convergent and stable solution \cite{ambati2015review}, and to the pertinent computational expense, often requiring substantial time and computational resources \cite{diehl2022comparative}.

One notable case of such a challenging coupled problem is phase-field fracture (PFF) modeling. Since the seminal work of Bourdin, Francfort and Marigo \cite{francfort1998revisiting, bourdin2000numerical}, the phase-field approach to simulate fracture has gained widespread popularity across the academic community. The PFF approach to fracture recasts the free-discontinuity problem posed by Griffith \cite{griffith1921vi} as an energy minimization formulation, regularizing the discrete crack through a finite-width, smeared representation. The latter is denoted by an auxiliary kinematic variable, the phase-field parameter ${\bf{\phi}}$, which ranges between 0 (intact material) and 1 (fully-damaged state). The phase-field approach offers a variational framework that circumvents the drawbacks of modeling discrete cracks using Linear Elastic Fracture Mechanics (LEFM). Over the past two decades PFF has been successfully applied in the investigation of several computationally challenging phenomena, including dynamic crack propagation \cite{ren2019explicit}, thermal cracks \cite{chu2017study, svolos2020thermal}, crack coalescence \cite{zhou2019propagation} and branching \cite{spatschek2011phase}, among others \cite{kristensen2020applications, bleyer2017dynamic, li2023review}. 

Despite its wide success, phase-field fracture modeling is also associated with formidable computational effort \cite{svolos2022fourth, schapira2023performance, zhuang2022phase}. This important drawback stems primarily from two reasons. First, the total potential energy functional is non-convex with respect to the kinematic variables \cite{ambati2015review}, the displacement field {\bf{u}} and the phase-field variable ${\bf{\phi}}$. Consequently, monolithic numerical solvers experience significant challenges to converge, and a considerable body of research has emerged to mitigate this phenomenon by developing more efficient solution schemes. The latter include modifications to the standard Newton-Raphson method (NR) \cite{wick2017modified}, quasi-Newton schemes \cite{kristensen2020phase}, staggered solvers \cite{hofacker2012continuum} limited-memory BFGS approaches \cite{jin2024novel}, energy-based arc-length techniques \cite{bharali2022robust}, advanced preconditioners \cite{li2024multiscale, badri2021preconditioning}, and domain decomposition-based methods \cite{svolos2020updating, rannou2024domain}. The second source of computational expense arises due to the need for a very fine discretization along the evolving crack path, which is necessary to accurately resolve the sharp gradients of the phase-field variable. Consequently, adaptive re-meshing techniques \cite{gupta2022adaptive, jin2024novel, yang2024finite, saberi2023multi, agrawal2021block, jaccon2023adaptive} have gained widespread popularity, dynamically refining the finite element mesh only where the crack is expected to propagate. Overall, the pursuit for more computationally efficient phase-field fracture simulations is a vibrant and actively explored area of research. 

Over the past few years, the advent of Machine Learning (ML) has fueled the development of several novel computational frameworks in the field of computational mechanics. Broadly, these methods can be classified into the following categories: data-driven, physics-informed, or a hybrid combination thereof. The reader is referred to \cite{herrmann2024deep} and \cite{jin2023recent} for comprehensive reviews on the application of machine learning-based methods in the field of computational mechanics. However, compared to other research areas such as hyperelasticity \cite{vlassis2020geometric, abueidda2022deep, mendizabal2020simulation} or plasticity \cite{huang2020machine, jang2021machine, fuhg2022machine}, ML-based frameworks for phase-field fracture are evidently more sparse. Below we provide an overview of the existing efforts, and we discuss the unique challenges associated with advancing such frameworks.

A major line of research involves frameworks dedicated to predicting the evolution stages of phase-field, including initiation, propagation, nucleation, and branching. To this end, early efforts involved the application of physics-informed neural networks (PINNs) as in \cite{goswami2020transfer}, with conceptually similar approaches being proposed in \cite{zheng2022physics} and \cite{manav2024phase}, all leveraging physics-based training paradigms. Along the same trajectory, frameworks based on more sophisticated operator architectures (DeepONets \cite{lu2021learning} and DeepOKANs \cite{abueidda2025deepokan}) were utilized in \cite{goswami2022physics} and \cite{kiyani2025predicting}. Moving beyond this scope, \cite{aldakheel2025physics} proposed a novel computational framework that embeds physics principles directly into the neural architecture; however, its application was demonstrated only for the one-dimensional problems. Finally, a constitutive modeling based approach originating from the physics-augmented neural networks (PANNs) \cite{klein2022finite} was recently applied to phase-field fracture \cite{dammass2025neural}. We note however that the latter approach is mostly grounded on hyperelastic modeling assumptions which often involve large-deformation considerations, a reliance which may render it less suitable for material systems with brittle or quasi-brittle responses.

Returning to first family of frameworks, it is important to highlight that, despite their promise, significant challenges remain that hinder their scalability and deployment in real-world, large-scale scenarios. First, regardless of the training strategy or network architecture, the offline learning task is still burdened by considerable computational overhead. Achieving accurate predictive behavior often necessitates one of the following: complex architectures with a large number of trainable parameters (such as operators), large number of training epochs in physics-constrained training, or excessive training datasets. Evidently, each of the aforementioned aspects is tied to prohibitively expensive training times for real-time problems. Additionally, given the path-dependent nature of crack propagation, these approaches rely either on repetitive re-training (transfer-learning) at every loading increment, or on fixed-length output predictions. The former quickly becomes impractical or even infeasible for large-scale simulations, while the latter limits the network applicability to predefined load incrementation schemes, again reducing the model generalizability across the time domain. Overall, there is a true need for new methods that combine the following virtues: they are \textit{physics-based}, \textit{computationally-efficient} and \textit{generalizable} across varying geometric and loading setups.

\subsection{The IFENN approach to multi-physics problems}

Acknowledging the limitations of purely FEM- and ML-based methods, hybrid modeling approaches have started gaining more attention \cite{thel2024introducing, mitusch2021hybrid, meethal2023finite, zhang2022simulation, aldakheel2025physics}. Along this trajectory, we proposed and have been continuously developing over the last several years the Integrated Finite Element Neural Network (IFENN) framework for non-local gradient damage and thermoelasticity coupled problems \cite{pantidis2023integrated, pantidis2023116160, pantidis2024fenn, abueidda2024variational, abueidda2024fenn}, which marks a conceptually distinct approach from the aforementioned frameworks. The core idea of IFENN is to decouple the governing equations into two solvers: a traditional FEM-based solver and a specialized neural network. Each solver targets a primary kinematic variable, and they exchange information at every iteration to ensure the seamless flow of information throughout the entire simulation. This hybrid approach delivers two key advantages: numerical robustness and accuracy that is guaranteed by minimizing the FEM residual, and computational efficiency that is achieved through the unmatched online predictive speed of neural networks. As a consequence, IFENN avoids the extreme computational cost of solving complex non-linear problems exclusively with FEM, while overcoming instabilities that are often associated with purely ML-based simulations. So far however, IFENN implementation has still been subject to some of the previously mentioned limitations: heavy and fixed-length \textit{Seq2Seq} models were used to capture temporal features, physics-based formulations suffered from high computational cost, and limited generalizability was observed in data-driven training setups. 

\subsection{Scope and Outline}
\label{Scope_and_Outline}

In this work, we present a new IFENN-based method for phase-field fracture modeling that achieves several fundamental milestones beyond the current literature of hybrid FEM-ML modeling. Specifically, the novel contributions of this paper are summarized as follows:

\begin{itemize}

    \item For the first time, a new IFENN scheme is proposed to model phase-field fracture propagation based on a physics-informed convolution neural network (PICNN). The adopted network is trained purely in an unsupervised context (without any labeled datasets), guided solely by the governing physics.

    \item Since the PICNN receives image-like input, we train the network on a pixel-based depiction of the phasefield-strain energy coupling. As a result, the proposed PICNN-IFENN setup eliminates the need for considering \textit{temporal features} in the network training stage. This time-independence has a twofold advantage: it drastically reduces the computational cost in the offline training and results in a network that can be seamlessly adapted to different load incrementation schemes in the online stage.

    \item We show that a highly efficient training approach with an ultra-small dataset is both feasible and effective. Specifically, the PICNN used in this study is trained on the strain energy density profiles from only two load increments, with a total training time of $\approx$ 5 minutes. 
    
    \item The utilized PICNN is agnostic to loading conditions and geometry variations, as long as the domain is rectangular and the mesh is structured. We show that the same PICNN can be readily implemented to model crack propagation across different geometries and at different loading/boundary conditions than those seen during training.  

    \item The resulting PICNN-IFENN solver is insensitive to the mesh size, and we show that computational savings scale with increasing mesh resolution. 
    
\end{itemize}

To encapsulate the above: we demonstrate that a single PICNN trained on only two load increments from a simple single-crack tension case, can - when embedded within IFENN - model phase-field propagation across a wide range of scenarios, including arbitrary rectangular domains, single and multiple interacting cracks, varying mesh densities, and arbitrary loading paths. The proposed formulation delivers radical breakthroughs that address many of the limitations in the existing methods and opens the door for many new possibilities in modeling coupled problems (especially phase-field fracture) using ML and coupled FEM-ML methods. In the following sections, we discuss in detail the formulation, functionality, and limitations of the proposed PICNN-IFENN method. 

The paper is structured as follows. Section \ref{Section:Phasefield_for_fracture} presents the basic theoretical aspects of the phase-field approach to fracture modeling. In Section \ref{Section:Methodology} we delve into the core of our methodology, and we present the structural and conceptual details of the proposed PICNN-IFENN approach for phase-field fracture. In Section \ref{Section:One_tine_training} we perform the one-time training of the proposed network, and the results from the numerical implementation are reported in Section \ref{Section:Results}. Finally, a discussion on our conclusions, current limitations, and ongoing work is provided in Section \ref{Section:Conclusions}.

\section{Phase-field approach to fracture}
\label{Section:Phasefield_for_fracture}

In this section, we present the theoretical basis of the phase-field fracture formulation for brittle materials. First, we introduce the fundamental concepts and then summarize the governing equations along with details of the numerical integration method.

\subsection{Formulation: Fundamental Concepts}
\begin{figure}[H]
    \centering
    \includegraphics[width=0.5\textwidth]{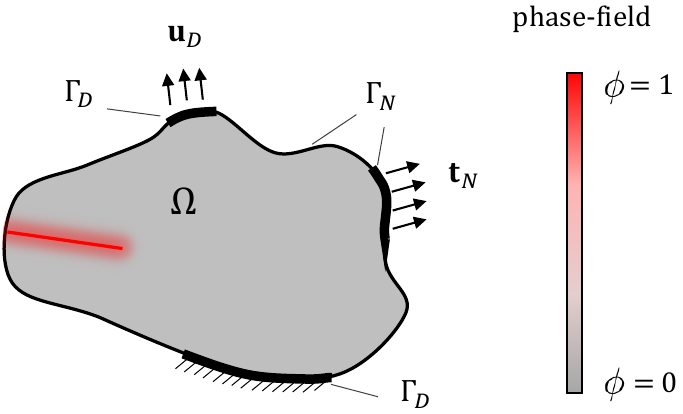}
    \caption{Schematic of a generic domain $\Omega$ with a sample crack that is represented by the phase-field variable $\phi$.}
    \label{Figure_schematic}
\end{figure}

We consider the domain of a brittle and deformable solid body, $\Omega \subset \mathbb{R}^\delta$, where $\delta \in \{1,2,3\}$ denotes its spatial dimension (as shown in Fig.~\ref{Figure_schematic}). The boundary, $\partial \Omega$, is divided into two disjoint subsets, $\Gamma_D$ and $\Gamma_N$, where Dirichlet and Neumann boundary conditions are imposed, respectively. The fracture topology is represented by a continuous (phase) field $\phi(\bm{x}, t): \Omega \times \mathbb{R}^+ \rightarrow [0,1] $, where $\phi = 0$ corresponds to the intact state and $\phi = 1$ to the fully damaged state at a material point $\bm{x} \in \Omega$ and time instant $t$.

Following the infinitesimal strain theory, the strain tensor is defined as the symmetric part of the displacement gradient, $\bm{\varepsilon}=\nabla^s \bm{u}:=\frac{1}{2} [ \nabla \bm{u} + (\nabla \bm{u})^\mathrm{T}]$, where $\bm{u}(\bm{x}, t): \Omega \times \mathbb{R}^+ \rightarrow \mathbb{R}^\delta$ represents the displacement field, $\nabla$ denotes the gradient operator, and the superscript $\mathrm{T}$ is the tensor transpose.

The phase-field approach to fracture was originally proposed by Bourdin~\emph{et al.}~\cite{bourdin2000numerical} as a generalization of the variational formulation of Francfort and Marigo \cite{francfort1998revisiting}, which in turn regards fracture as a competition between bulk elastic energy and crack surface energy \cite{griffith1921vi}. Adopting the AT2 model \cite{pham2011gradient}, the deformation of the body and the crack topology are determined by minimizing the following total energy functional 
\begin{equation} \label{eq:EnergyFunctional}
    \mathcal{E}(\bm{u}, \phi) = \int_\Omega \Psi(\bm{\varepsilon}, \phi) d\bm{x} + \int_\Omega \frac{G_c}{2 l_c} \left( \phi^2 + l_c^2 \Vert \nabla \phi \Vert^2 \right) d\bm{x} - \int_\Omega \bm{b} \cdot \bm{u} d\bm{x} - \int_{\Gamma_N} \bar{\bm{t}} \cdot \bm{u} d\bm{s} \, ,
\end{equation}
where $\Psi$ is the \emph{degraded} strain energy density (reduced due to material damage), $G_c$ denotes the critical energy release rate, $\bm{b}$ represents the body forces, and $\bar{\bm{t}}$ is the traction vector applied on the Neumann boundary $\Gamma_N$.

In Eqn.~\ref{eq:EnergyFunctional}, the first term corresponds to the elastic energy and the second term to the regularized crack surface energy (or fracture energy). Regularization is achieved through the length-scale parameter $l_c$ that regulates the thickness of the smeared crack representation.   

Under the assumption of linear elasticity, the \emph{undamaged} strain energy density is given by
\begin{equation}
    \Psi_0(\bm{\varepsilon})=\Psi(\bm{\varepsilon},\phi=0) = \frac{\lambda}{2} \left[ \operatorname{tr}(\bm{\varepsilon}) \right]^2 + \mu \, \bm{\varepsilon} : \bm{\varepsilon} \, ,
\end{equation}
where $\lambda$ and $\mu$ are the Lam\'e parameters, `$\operatorname{tr}$' represents the trace operator for tensors, and `:' denotes the double contraction between two second-order tensors. The elastic strain energy density is additively split into tensile and compressive parts, $\Psi^+$ and $\Psi^-$, respectively. Similarly to \cite{miehe2010thermodynamically}, we adopt the spectral decomposition of the strain tensor as follows
\begin{equation}
    \bm{\varepsilon} = \sum_{i=1}^{\delta} \varepsilon_i \, \bm{n}_i \otimes \bm{n}_i \, ,
\end{equation}
where $\{ \varepsilon_i \}_{i=1,\dots,\delta}$ denote the principal strains and $\{ \bm{n}_i \}_{i=1,\dots,\delta}$ are the corresponding principal directions. The tensile/compressive components of $\Psi_0$ are defined by
\begin{equation}
    \Psi_0^\pm (\bm{\varepsilon}) = \frac{\lambda}{2} \langle \operatorname{tr}(\bm{\varepsilon}) \rangle_\pm^2 + \mu \, \bm{\varepsilon}_\pm : \bm{\varepsilon}_\pm \, ,
\end{equation}
where the Macaulay brackets are defined by $\langle x \rangle_\pm = (x \pm \vert x \vert )/2$ and the positive/negative components of the strain tensor read
\begin{equation}
    \bm{\varepsilon}_\pm = \sum_{i=1}^{\delta} \langle \varepsilon_i \rangle_\pm \, \bm{n}_i \otimes \bm{n}_i \, .
\end{equation}
Finally, the degraded elastic strain energy density can be expressed as
\begin{equation}
    \Psi(\bm{\varepsilon}, \phi) = g(\phi) \Psi^+_0(\bm{\varepsilon}) + \Psi^-_0(\bm{\varepsilon}) \, ,
\end{equation}
where $g(\phi)$ is a degradation function that controls the elastic energy reduction due to damage. It is noteworthy that only $\Psi^+$ is multiplied by the degradation function, in order to prevent crack initiation in compression-dominated regimes. Interested readers are referred to \cite{ambati2015review,van2020strain,vicentini2024energy} for alternative energy decompositions and to \cite{kuhn2015degradation,sargado2018high,svolos2023convexity} for various degradation functions.




\subsection{Governing equations and their numerical integration}

By minimizing the total energy functional as expressed in Eqn.~\ref{eq:EnergyFunctional}, the Euler-Lagrange equations of the model can be derived. In this work, we adopt the hybrid isotropic/anisotropic formulation presented in \cite{ambati2015review}, where the governing equations of the problem is given by
\begin{equation} \label{eq:GoverningEquations} \left.
\begin{aligned} 
    \nabla \cdot \bm{\sigma} + \bm{b} & = \bm{0} \\
     - G_c \, l_c \Delta \phi + \frac{G_c}{l_c} \phi + g^\prime(\phi) \Psi^+_0  & = 0  \\ 
\end{aligned} \right\} \, .
\end{equation}
Here $\nabla \cdot$ represents the divergence operator, and $\Delta$ denotes the Laplacian. In this work, the degradation function is defined as $g(\phi)=(1-\phi)^2$, without loss of generality for the proposed framework. The Cauchy stress tensor, $\bm{\sigma}$, is expressed as follows
\begin{equation}
    \bm{\sigma} = g(\phi) \frac{\partial \Psi_0}{\partial \bm{\varepsilon}} = g(\phi) \left[ \lambda \operatorname{tr}(\bm{\varepsilon}) \bm{I} + 2 \mu \bm{\varepsilon} \right] \, ,
\end{equation}
where $\bm{I}$ denotes the second-order identity tensor. Given pseudo-time-dependent prescribed displacements $\bar{\bm{u}}$ and traction-free external surfaces, the boundary conditions are defined as follows
\begin{equation} \left.
\begin{aligned}
    \bm{u}(\bm{x},t) = \bar{\bm{u}}(\bm{x}, t) \quad \text{on} \quad \bm{x} \in \Gamma_D \\
    \bm{\sigma} \bm{n}(\bm{x})= \bar{\bm{t}}=\bm{0} \quad \text{on} \quad \bm{x} \in \Gamma_N \\
    \nabla \phi \cdot \bm{n}(\bm{x}) = 0 \quad \text{on} \quad  \bm{x} \in \partial \Omega
\end{aligned} \right\} \, ,
\end{equation}
where $\bm{n}(\bm{x})$ denotes the unit outward normal vector to the boundary at the point $\bm{x} \in \partial \Omega$.

After discretizing the pseudo-time $t$ into time steps, denoted by the set $\{t_n\}_{n=0,\dots,N_T}$, where $t_0=0$ and $N_T$ is the total number of time increments, we then introduce a history variable
\begin{equation}
    H_n = \max_{t_0, \dots, t_{n-1}} \Psi_0^+ \, ,
\end{equation}
which replaces the crack-driving force $\Psi_0^+$ in Eqn.~(\ref{eq:GoverningEquations})$_2$, in order to enforce crack irreversibility. Finally, we adopt the finite element method (FEM) along with a staggered solver \cite{miehe2010phase} using the FEniCS computing platform \cite{baratta2023dolfinx,alnaes2014unified} to numerically integrate the governing equations.


\section{Methodology}
\label{Section:Methodology}

In this section, we outline the key components of the PICNN-based IFENN approach for modeling phase-field fracture. Section \ref{Section:IFENN_scheme} reviews the fundamentals of the IFENN solver for multi-physics problems. Section \ref{Section:PICNN_for_phasefield} discusses the design and training details of the neural network of choice; in this work, it is a physics-informed convolutional network. Finally, Section \ref{Section:IFENN_with_PICNNs} presents the integrated workflow and implementation details of the PICNN-based IFENN scheme. 

\subsection{IFENN for multi-physics problems}
\label{Section:IFENN_scheme}

\begin{figure}[t!]
    \centering
    \includegraphics[width=1\textwidth]{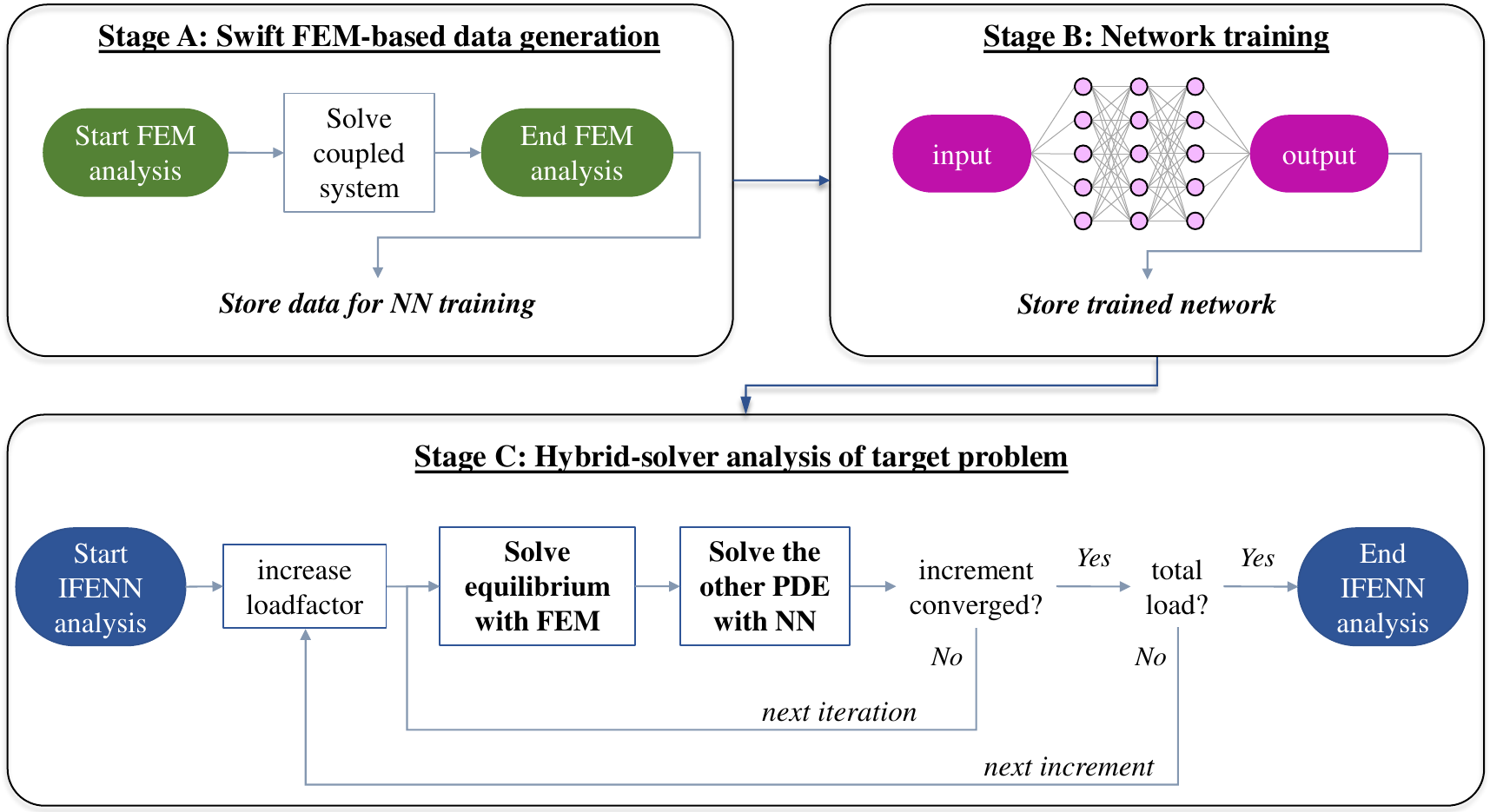}
    \caption{IFENN workflow. {\bf{a.}} Stage A: a swift FEM analysis on a simplified version of the target problem is performed, to generate the training data {\bf{b.}} Stage B: a problem-specific neural network is selected and trained {\bf{c.}} Stage C: the computationally expensive target problem is solved with the hybrid solver.}
    \label{Figure_IFENN}
\end{figure}

IFENN is designed to accelerate the numerical simulation of multi-physics problems. An overview of the framework is presented in Fig. \ref{Figure_IFENN}, where its three main components are outlined. The core idea of the method is to solve a computationally expensive problem by splitting the governing equations in two solvers: mechanical equilibrium is solved through a conventional FEM solver, while a specialized pre-trained neural network computes the other physics-related PDE. A detailed explanation of each stage is provided below.

In the first stage, shown in Fig. \ref{Figure_IFENN}a, an FEM analysis of a simplified version of the target problem is performed. This step is intentionally kept computationally inexpensive, to ensure that the overall acceleration benefit of the framework is preserved. For example, this analysis could be performed on a coarse-mesh discretization of the domain, as was done for non-local gradient damage problems in \cite{pantidis2024fenn}. The objective of this stage is to generate the training data that are necessary in the subsequent network training. The nature and type of the data depends on the problem under consideration. For instance, for a physics-based training of non-local damage, the data would include Gauss-point coordinates and the local equivalent strain profile \cite{pantidis2023integrated, pantidis2023116160}, whereas for a data-driven training variant the labeled values of non-local strain would also be necessary \cite{pantidis2024fenn}.

The second stage of IFENN, presented in Fig. \ref{Figure_IFENN}b, involves the design and training of a specialized neural network to learn the governing PDE. The selection of the network architecture should be guided by the nature of the target problem. For example, previous implementations of IFENN have employed fully-connected multi-layer perceptrons (MLPs) and temporal convolutional networks (TCNs), while in this work we explore the utilization of Convolutional Neural Networks (CNNs). We note that depending on the learning objective, the network can be data-driven, physics-informed, or trained with a combination of both methods.

Finally, in the third stage which is shown in Fig. \ref{Figure_IFENN}c, the target problem is solved with a hybrid solver that integrates the trained neural network within the non-linear FEM loop. The two solvers operate in a staggered fashion, being alternately called within each load increment until convergence is achieved. Computational gains stem on the premise that the combined computational expense from all the stages above is smaller than the cost of the FEM-only analysis of the target coupled problem.

\subsection{Physics Informed Convolutional Neural Networks (PICNNs) for phase-field fracture}
\label{Section:PICNN_for_phasefield}

\subsubsection{Why PICNNs?}
\label{Section:Why_PICNNs}

In this study we aim to learn the underlying physical relationship between the strain energy density $H$ and phase-field $\phi$ {\bf{without incorporating any temporal dependencies}}. Our approach is driven by two key motivations: a) reducing the overhead associated with sequence models, which includes the costly data acquisition and intensive training, and b) arriving at a network that is agnostic to time incrementation schemes, thus avoiding the constraints imposed by the fixed-length nature of sequence models. As such, we depart from the conventional idealization of $H$ and $\phi$ as time-sequences and we refrain from utilizing sequence models in the online IFENN stage. 

Instead, we propose that it is sufficient to learn the $H \rightarrow \phi$ mapping locally, only at the fracture process zone, and that this information can be utilized at any load increment along the advancing crack front. This can be achieved by adopting a geometry-based training approach and by learning the spatial features of the $H$ and $\phi$ distributions. Our implementation leverages the powerful pattern recognition capabilities of CNNs, employing a physics-informed variant (PICNNs) that is trained exclusively on physical principles.

CNNs are a class of deep learning models that are specifically developed to process data with a grid-like underlying structure. Their architecture is built around convolutional layers, which apply learnable \textit{filters} - also termed \textit{kernels} - across the input domain to automatically detect local patterns. This mechanism allows CNNs to efficiently extract spatially hierarchical features, such as edges and shapes, and they can then be used to recognize similar patterns in unseen testing datasets. While the primary application field of CNNs has traditionally been image processing \cite{archana2024deep}, they have been successfully applied in other tasks such as audio processing and speech recognition \cite{nanni2021ensemble, alsobhani2021speech}.

\subsubsection{Gauss point-based to pixel-based representation}
\label{Section:GP_to_pixels}

\begin{figure}[b]
    \centering
    \includegraphics[width=0.9\textwidth]{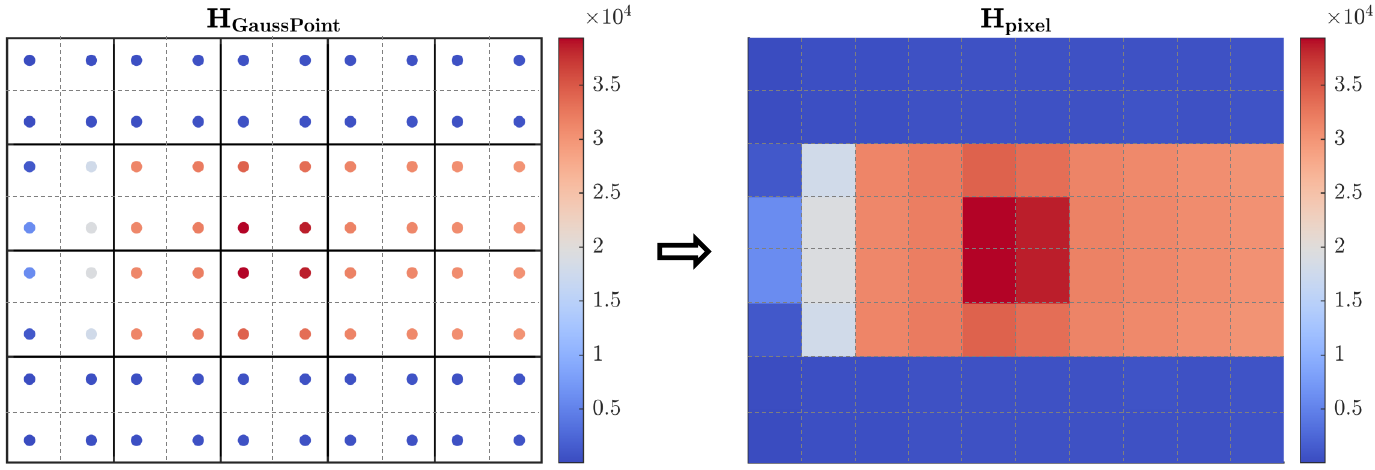}
    \caption{Schematic transformation of the strain energy density profile evaluated at the Gauss points to an equivalent pixel-based representation.}
    \label{Figure_GPtoPixels}
\end{figure} 

The objective of the PICNN is to learn mappings between the strain energy density $H$ (input variable) and the phase-field variable $\phi$ (output variable), with an emphasis on the behavior at the crack front. The information for both variables is defined at the Gauss (integration) points, henceforth denoted as GPs. To make the input compatible with the architectural requirements of CNNs, the $H$-profile must first be converted into a pixel-based representation. The latter is a 2D matrix of pixel values, where each pixel encodes the average $H$ over its area of influence. This transformation ensures that the information from the original $H$-field is preserved, while enabling the PICNN training.

In this context, a straightforward way of implementation is to consider rectangular domains which are discretized with first-order square finite elements. The combination of both features results in a very close alignment between the locations of the Gauss points and the pixel centers, as schematically depicted in Fig. \ref{Figure_GPtoPixels}. Evidently, this match is not exact, as the Gauss points are slightly shifted compared to the assumed center of each pixel. However, the discrepancy is assumed to be negligible, and - very importantly - this approach enables a direct, one-to-one mapping between the GP-based values and the pixel-based values. Also, this arrangement eliminates the need for explicitly encoding the GP coordinates in the training, as the pixel indices inherently capture the spatial arrangement of the $H$-profile. At the end, the input of the PICNN is a single 3D tensor of the $H$ values with size: $N \times P_{x} \times P_{y}$ where $N$ is the number of load increments and $P_{x,y}$ are the number of pixels in the x- and y-direction respectively. An evident limitation of this setup is the need to use the same mesh discretization across the entire domain, however this can be remedied using well-established projection and spatial-interpolation methods \cite{lam1983spatial, minion1996projection, teegavarapu2012geo}.

\subsubsection{PICNN architecture}
\label{Section:PICNN_architecture}

\begin{figure}[t!]
    \centering
    \includegraphics[width=1\textwidth]{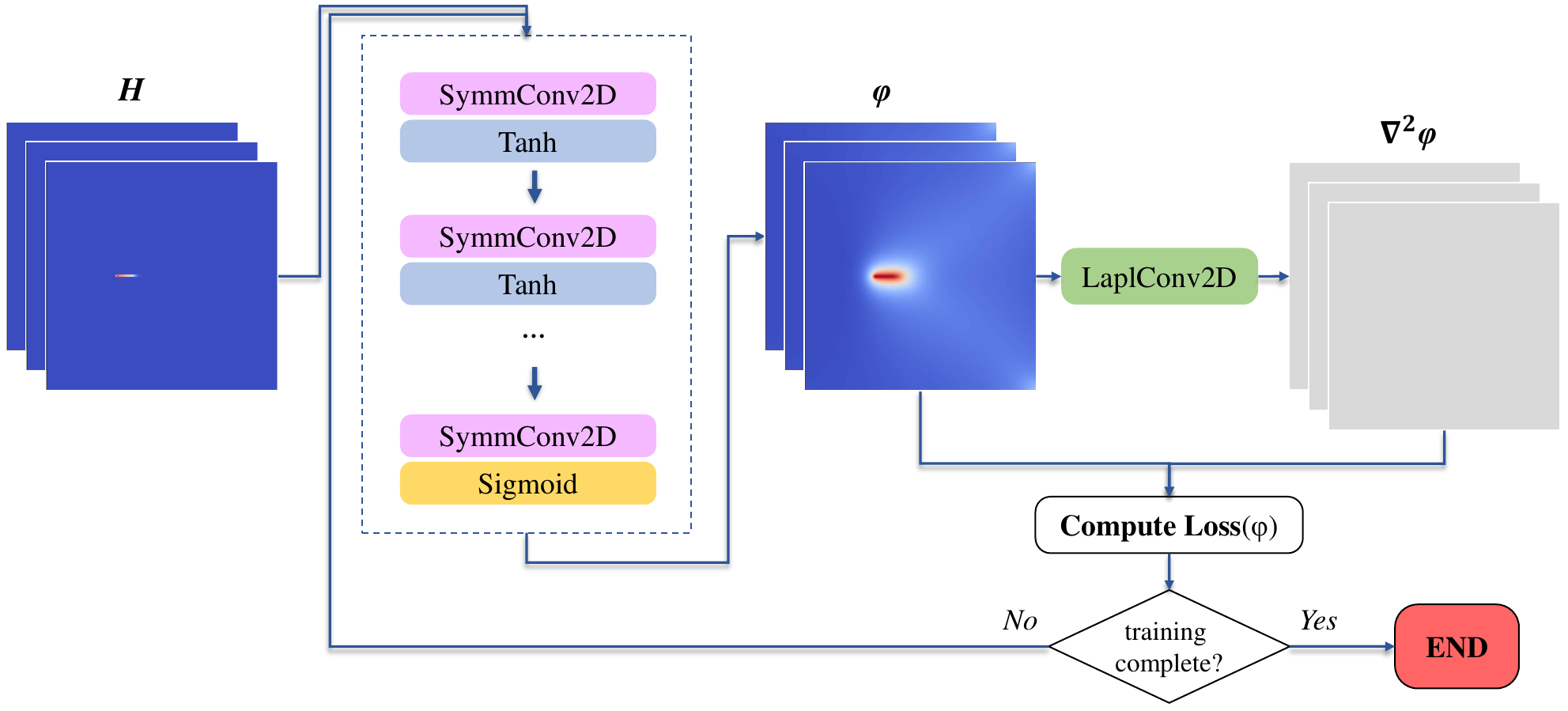}
    \caption{PICNN architecture and training process}
    \label{Figure_PICNN_training}
\end{figure}

A schematic of the proposed network architecture is shown in Fig. \ref{Figure_PICNN_training}. The input and output of the network are both image-like data. The PICNN consists of a series of \emph{symmetric} convolution layers followed by the hyperbolic tangent function. The last convolution layer is followed by the sigmoid function, which ensures that the predicted $\phi$ values are constrained between 0 and 1. The phase-field Laplacian is then computed using a predefined convolution filter. This step, along with the formulation of the loss function, are detailed in the next subsection. We emphasize the simplicity of the proposed architecture, which does not include any batch-normalization or pooling operations. Also, we deliberately omit the typical "flatten" operation, which is a common constituent of the CNN setup. This choice allows the predicted $\phi$-profile to retain the same structure and size as the input $H$-profile.

Next, we underline a critical structural modification that we implement in our setup compared to the vanilla-type CNN layers. Since our goal is to deploy the same PICNN across varying load scenarios, it is critical that the learned model remains insensitive to the specific orientation of the $H$-profile. Essentially, the trained network must be capable of predicting the correct phase-field regardless of the directionality of the $H$-input. To this end, we directly embed translational and rotational invariance in our design by enforcing $symmetry \ by \ construction$ into the kernels of each layer. Each layer uses 5 $\times$ 5 kernels. Instead of allowing all 25 entries to be independently learned, we impose a symmetry condition along both the horizontal and vertical axis. As a result, only 6 parameters are uniquely defined and learned per kernel, while the symmetry conditions determine the remaining entries. A schematic view of the designed kernels is shown in Fig. \ref{Figure_PICNN_details}a. 

\subsubsection{PICNN training}
\label{Section:PICNN_training}

In order to construct the PICNN loss function, we first need to compute the Laplacian of $\phi$ at every pixel. We approach this task using the finite difference method, and below we elaborate on our approach which is inspired by \cite{zhao2023physics}. 

Let us consider a square computation domain $\Omega$ with side length $l$. The domain is discretized into $n \times n$ elements with a step of $h = l/n$. These elements are essentially the pixels, whose centers (denoted with yellow in Fig. \ref{Figure_PICNN_details}b) are also spaced by $h$ in the cartesian coordinate system. The pixel centers reside on a structured grid, whose indices are denoted as $i$ and $j$ along the x and y directions, respectively. Using the finite difference method, the five-point formula for the second-order spatial derivative of $\phi$ in the x-direction yields: 

\begin{figure}[t!]
    \centering
    \includegraphics[width=0.65\textwidth]{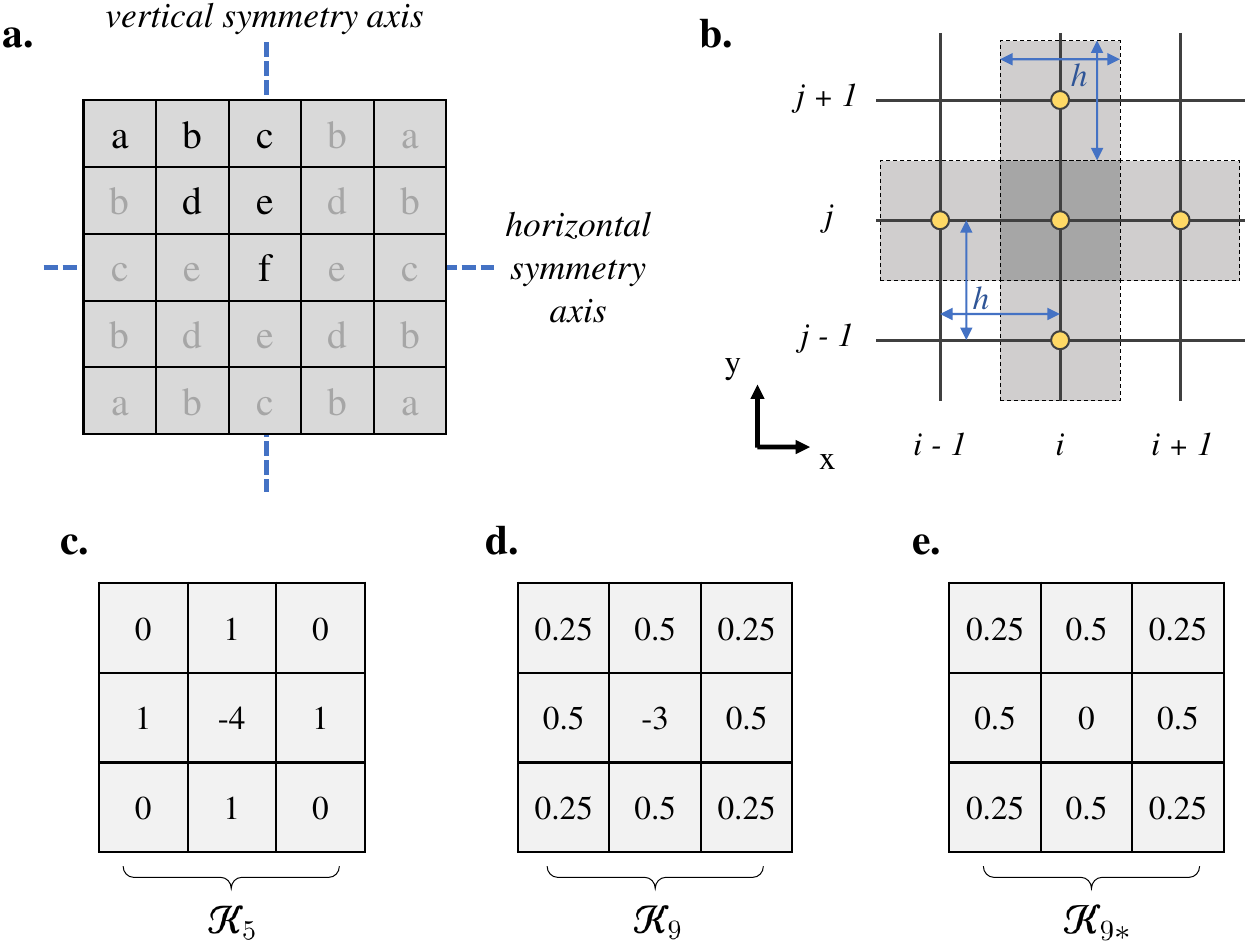}
    \caption{{\bf{a.}} Double-symmetric arrangement of $5\times5$ kernels used at each layer, resulting in 6 independent trainable entries per kernel. {\bf{b.}} Schematic grid of pixels and the finite difference approximation using their centers as reference points. {\bf{c.}} Laplacian convolution filters for 5-point stencil (nominal), 9-point stencil (nominal), and the modified 9-point stencil.}
    \label{Figure_PICNN_details}
\end{figure} 

\begin{equation}
    \frac{\partial^2 \phi(x_{i},y_{j})}{\partial x^2} = \frac{1}{h^2}\left[ \phi(x_i + h, y_j) - 2 \phi(x_i, y_j) + \phi(x_i - h, y_j)\right] - \frac{h^2}{24} \left[ \frac{\partial^4 \phi(\xi_1, y_j)}{\partial x^4} + \frac{\partial^4 \phi(\xi_2, y_j)}{\partial x^4}  \right]
\label{Eqn:phi2_x2}
\end{equation}

\noindent where $x_i - h \leq \xi_1, \xi_2 \leq x_i + h$. 

Similarly, the expression for the second derivative in the y-direction reads:

\begin{equation}
    \frac{\partial^2 \phi(x_{i},y_{j})}{\partial y^2} = \frac{1}{h^2}\left[ \phi(x_i, y_j + h) - 2 \phi(x_i, y_j) + \phi(x_i, y_j - h)\right] - \frac{h^2}{24} \left[ \frac{\partial^4 \phi(x_i, \eta_1)}{\partial y^4} + \frac{\partial^4 \phi(x_i, \eta_2)}{\partial y^4}  \right]
\label{Eqn:phi2_y2}
\end{equation}

\noindent where $y_j - h \leq \eta_1, \eta_2 \leq y_j + h$. 

We henceforth denote ($x_{i} - h), (x_{i} + h$) as ($x_{i-1}), (x_{i+1}$) and similar for the other direction. The Laplacian of the phase-field variable is then formed by summing Eqns. \ref{Eqn:phi2_x2} and \ref{Eqn:phi2_y2}: 

\begin{equation}
    \nabla^{2}\phi(x_{i},y_{j}) = \dfrac{\partial^2 \phi}{\partial x^2} + \dfrac{\partial^2 \phi}{\partial y^2} = \dfrac{1}{h^2} \underbrace {\left[ \phi(x_{i+1},y_{i}) + \phi(x_{i-1},y_{i}) + \phi(x_{i},y_{i+1}) + \phi(x_{i},y_{i-1}) - 4\phi(x_{i},y_{i}) \right]}_{\mathbcal{K_{5}}} + O(h^{2})
\end{equation}

In the expression above, $O(h^{2})$ is the truncation error and it is henceforth omitted from the computations. Also, the expression $\left[ \phi(x_{i+1},y_{i}) + \phi(x_{i-1},y_{i}) + \phi(x_{i},y_{i+1}) + \phi(x_{i},y_{i-1}) - 4\phi(x_{i},y_{i}) \right]$ is denoted for brevity as $\mathbcal{K_{5}}(\phi)$, which is simply the Laplacian convolution kernel based on the 5-point stencil, shown in Fig. \ref{Figure_PICNN_details}c. A similar kernel, denoted as $\mathbcal{K_{9}}(\phi)$ and shown in Fig. \ref{Figure_PICNN_details}d, can be derived following the 9-point formulation. In our training we adopt the latter, with a minor implementation modification. Specifically, we find that the training is conducted more efficiently when the Laplacian term is computed as:

\begin{equation}
    \nabla^{2}\phi(x_{i},y_{j}) = \frac{1}{h^{2}} \left[ \mathbcal{K_{9*}}(\phi(x_{i},y_{j})) - 3\phi(x_{i},y_{j}) \right]
\end{equation}

\noindent where the filter $\mathbcal{K_{9*}}$ is shown in Fig. \ref{Figure_PICNN_details}e. This implementation aligns in spirit with the observations reported in \cite{zhao2023physics}. Ultimately, the residual of the phase-field PDE is expressed as:

\begin{equation}
    \mathbcal{R}(x_{i},y_{j}) = \dfrac{G_{c}}{l_{c}} \phi(x_{i},y_{j}) - G_{c}l_{c} \nabla^{2}\phi(x_{i},y_{j}) - 2(1-\phi(x_{i},y_{j}))H(x_{i},y_{j})
\end{equation}

\noindent where $H$, $\phi$ and $\nabla^{2}\phi$ are defined each each pixel $(x_{i},y_{j})$. The loss function is then formulated as the $2^{nd}$ norm of the PDE residual:

\begin{equation}
    \mathbcal{L} = ||\mathbcal{R}||_{2}
\label{Eqn:PICNN_loss}
\end{equation}

The network is trained to minimize the loss function of Eqn. \ref{Eqn:PICNN_loss} using the Adam optimizer for a predefined number of epochs. Our implementation is built on $pytorch$.

\subsection{IFENN with PICNNs: implementation of a general framework for phase-field fracture}
\label{Section:IFENN_with_PICNNs}

\begin{figure}[t]
    \centering
    \includegraphics[width=0.75\textwidth]{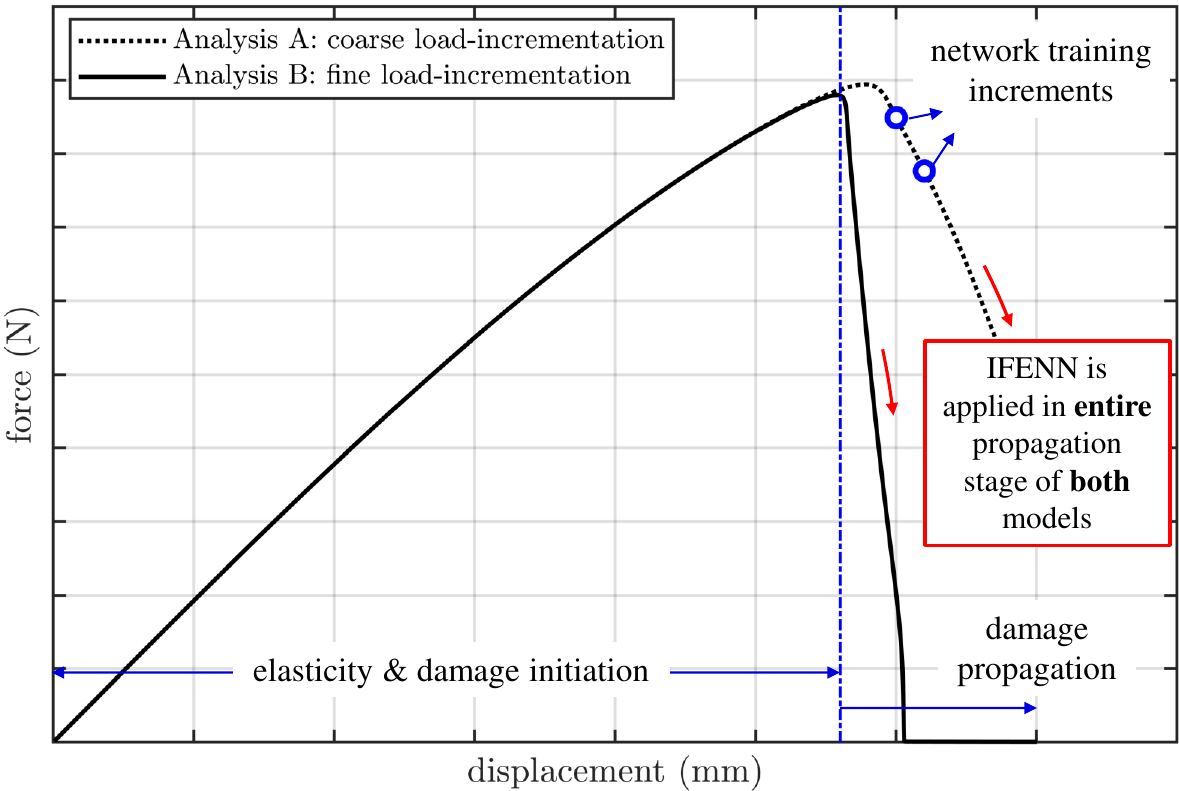}
    \caption{Schematic overview of PICNN-IFENN for phase-field fracture. The PICNN is trained on an ultra-low amount of propagation increments from a fast simulation. It is then used within IFENN to simulate the entire propagation phase of phase-field, regardless of the loading step scheme.}
    \label{Figure_IFENN_PFF}
\end{figure}

In this section we demonstrate the overarching workflow of the PICNN-IFENN framework for phase-field fracture. A schematic of our approach is shown in Fig. \ref{Figure_IFENN_PFF}. This graph presents two force-displacement curves for the same sample problem. The dashed line corresponds to Analysis A, a cost-effective simulation with a coarse load incrementation scheme (larger loading step), which is used to generate our training $H$-profiles. The solid line represents Analysis B, a more accurate but expensive simulation (smaller loading step), which is the target solution that we aim to reproduce with IFENN. Regardless of the loading step size, the phenomenon is characterized by two distinct stages: an initial phase of linear elasticity and damage initiation, which is then followed by damage propagation. To train the PICNN, we extract a minimal number of $H$-profiles from the propagation stage of Analysis A. Then, as an intermediate sanity check, we re-conduct Analysis A with conventional FEM until the damage propagation commences, and at this point we switch to IFENN and analyze the propagation stage with the hybrid solver. The final step (which is also our main goal) is to reproduce Analysis B in the same fashion: we solve the problem with FEM until the end of damage initiation, and utilize IFENN for the entire propagation stage. 

In order to execute the IFENN simulation, we perform within each load increment the following steps: 

\begin{itemize}

    \item Step 1: Solve the equilibrium equation. This step updates the nodal values of the displacement as well as the GP-values of $H$. We note that the solution to this equation also depends on the values of $\phi$ from the previous load increment.
    
    \item Step 2: Convert $H$ from the GP-level to the pixel-based representation. This is necessary to conform with the format that is expected by the PICNN. 

    \item Step 3: Predict the $\phi$-profile using the trained PICNN (online inference). This step leads to the new pixel-based idealization of the phase-field variable. 

    \item Step 4: Project the pixel-based values of $\phi$ back to each Gauss point. This step is necessary to update the phase-field values correctly at each finite element.

    \item Step 5: Check for convergence. If convergence is not achieved then return to Step 1 and repeat the above, otherwise proceed to the next load increment.
    
\end{itemize}

The workflow above reveals the integrative nature of IFENN: both an FEM-based and a NN-based PDE solver are interchangeably executed within a non-linear loop. The output of each solver becomes the input of the other, and this process continues seamlessly until the convergence criterion is satisfied. Finally, extensive preliminary analyses revealed that results improve significantly when the following implementation details are applied:

\begin{itemize}

    \item Though the strain energy density is freely updated at every iteration, its profile is capped between 0 and a maximum value (i.e. $5 \times 10^{4}$ or $1 \times 10^{5}$) when passed as input into the PICNN. This step ensures that the numerical values of $H$ during inference remain within the same range encountered by the PICNN during training, which in turn improves the PICNN online performance.

    \item Irreversibility conditions are imposed to both $H$ and $\phi$. While in a typical FEM-only setup one condition should suffice, we find that the PICNN does not always guarantee a non-decreasing $\phi$ in response to a non-decreasing $H$. This behavior is actually expected given the nature of the convolution operations involved. Therefore, to ensure physically consistent predictions, we impose irreversibility on $\phi$ as well. 

    \item We observed that applying a Gaussian-based smoothing filter on the predicted $\phi$-profile generally enhances the quality of the results. This step is motivated by the tendency of the PICNN to produce slightly narrower crack widths. A subsequent diffusion-based correction effectively mitigates this issue.  
    
\end{itemize}

\section{One-time PICNN training}
\label{Section:One_tine_training}

\begin{figure}[H]
    \centering
    \includegraphics[width=0.35\textwidth]{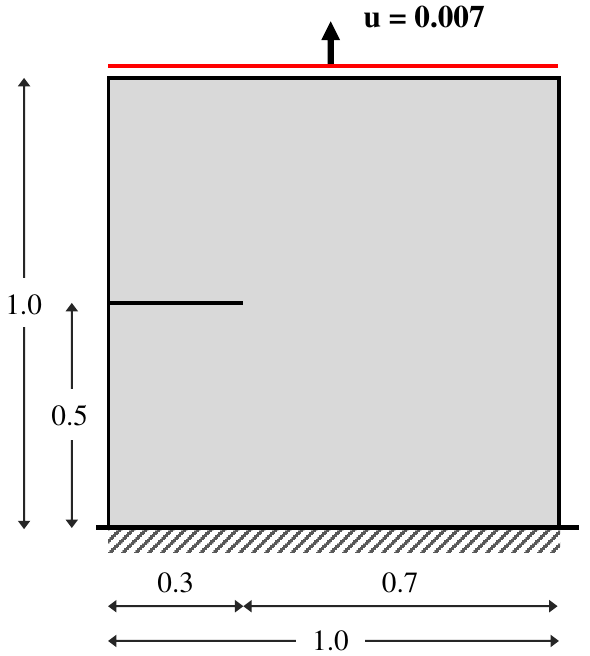}
    \caption{Geometric and loading/boundary condition details for the single-notch tension (SNT) problem}
    \label{Figure_SNT_schematic}
\end{figure}

In this section we present the details of the PICNN training stage, and we emphasize that the same PICNN will be used across all the numerical examples in the next section. The PICNN is trained on $H$-profiles generated from a single-notch specimen under tensile loading (SNT) model. The geometry and loading-boundary conditions of the SNT problem are shown in Fig. \ref{Figure_SNT_schematic}. This is a square domain of side length $l_{x} = l_{y}$ = 1.0 mm, with a horizontal left-side notch of length $l_{n}$ = 0.3 mm at the middle. The domain is fixed along the bottom edge and a tensile displacement is applied at the top. The material parameters are: first Lam\'e constant $\lambda =$ 121154 N/mm$^{2}$, second Lam\'e constant $\mu$ = 80770 N/mm$^{2}$, critical energy release rate $G_{c}$ = 2.7 N/mm, and characteristic length $l_{c}$ = 0.03 mm. The domain is discretized with a uniform square mesh using 40000 elements. The resulting model has a characteristic-to-element length ratio $l_{c} / l_{elem}$ = 6.0 and it is termed \textit{SNT1}. The loading is imposed in a monotonic displacement-driven context. In order to generate the dataset for the PICNN training we use a deliberately coarse incrementation, which is not intended to reproduce the well-documented brittle response of this specimen \cite{ambati2015review, miehe2010phase}. Instead, it only serves as a rapid preliminary analysis from which only two load increments for the PICNN training are extracted. This load case is denoted as \textit{I} and the displacement is incremented by $\Delta u = 2 \times 10^{-5}$ mm until a value of $u_{max}$ = 0.007 mm, resulting in 350 equidistant load increments. Within each increment the governing equations are solved using a single-pass (one staggered iteration) scheme, first solving the mechanical equilibrium, followed by the phase-field equation.

\begin{figure}[b!]
    \centering
    \includegraphics[width=0.7\textwidth]{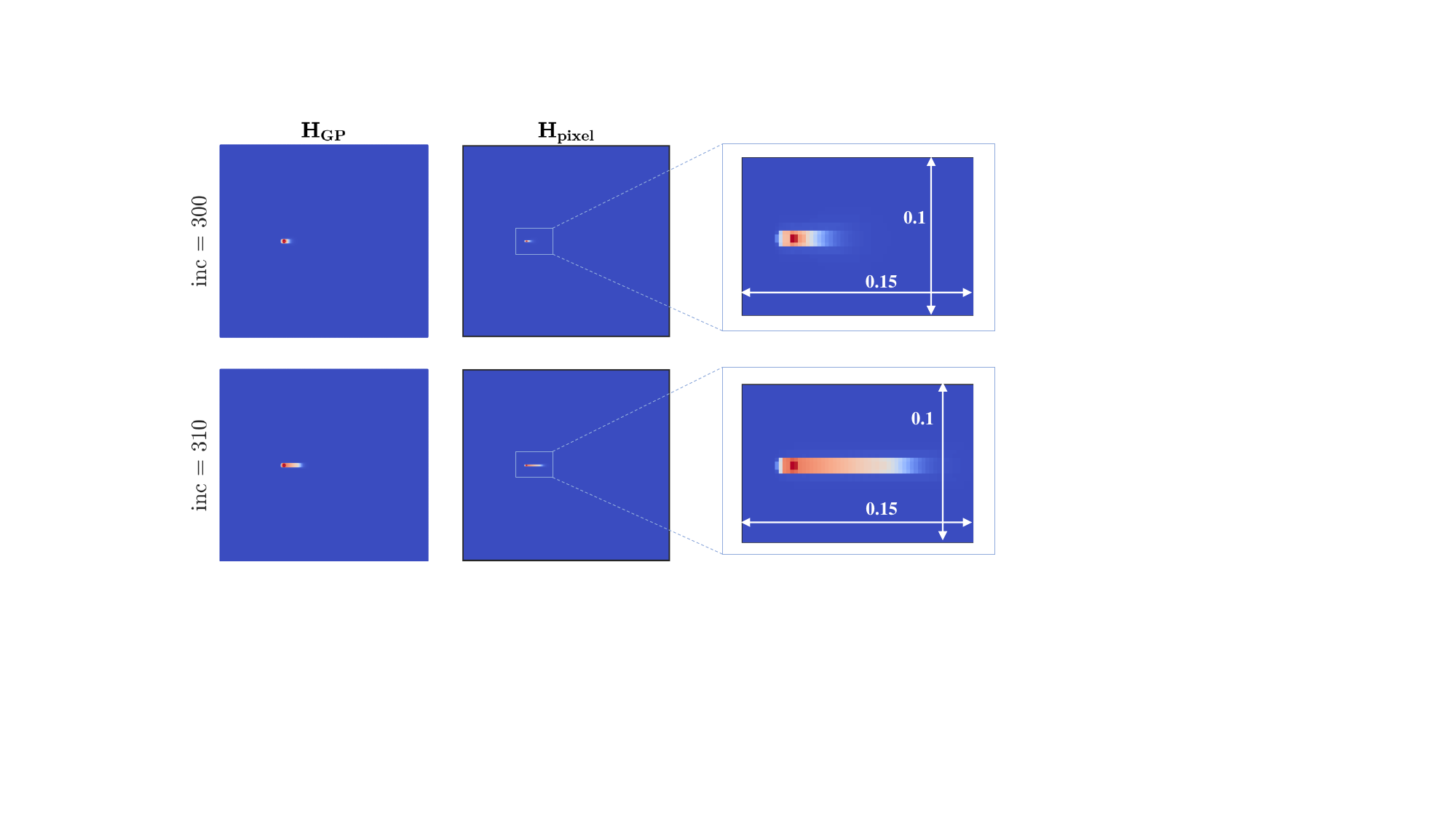}
    \caption{Strain energy density profiles of the two load increments used in training. The graphs on the left depict Gauss point-based distributions, which are then converted to pixel-based maps as training input for the PICNN.}
    \label{Figure_PICNN_Hprofiles}
\end{figure} 

\begin{figure}[t!]
    \centering
    \includegraphics[width=1\textwidth]{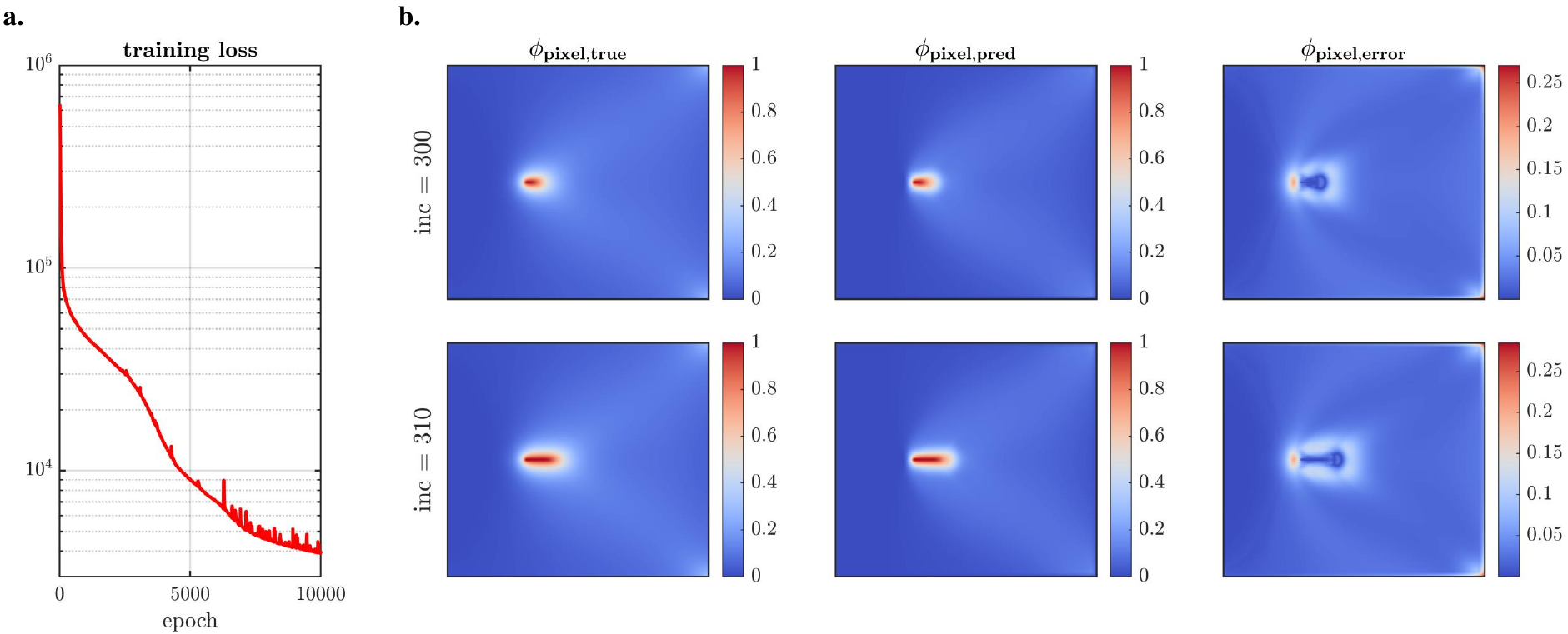}
    \caption{{\bf{a.}} Evolution of PICNN training loss function {\bf{b.}} True, predicted and absolute error $\phi$-maps for the two load increments used in training.}
    \label{Figure_PICNN_phiprofiles}
\end{figure}

The PICNN consists of 4 symmetric convolutional layers: the first three are followed by the hyperbolic tangent function, and the final layer is followed by a sigmoid function to constrain the phase-field output within the range [0-1]. Each layer has 24 input and output channels, with the exception of the first layer's input and last layer's output being 1 channel. As discussed in Section \ref{Section:PICNN_architecture}, the kernels of each layer are $5 \times 5$ in size and exhibit double symmetry, resulting in six independent, trainable entries per kernel. The PICNN is trained on the strain energy density profiles at two load increments (inc = [300, 310]) from the SNT1-I FEM analysis. These profiles are presented in Fig. \ref{Figure_PICNN_Hprofiles}a. It is crucial to highlight the extreme spatial localization of the H-profiles, which underscores the inherent difficulty of the learning task. As previously discussed, the Gauss point-based values are first converted to an equivalent pixel-based map to facilitate the training. The network is trained for 10000 epochs using the Adam optimizer with a learning rate $lr = 10^{-4}$. We highlight that the total training time takes approximately 5 minutes on a GPU-enabled DELL laptop. This is an exceptionally fast training process that is orders of magnitude faster than commonly reported in the literature \cite{manav2024phase, goswami2022physics, goswami2020transfer, kiyani2025predicting}. The minimization of the loss function is shown in Fig \ref{Figure_PICNN_phiprofiles}a, while Fig. \ref{Figure_PICNN_phiprofiles}b depicts the true, predicted and absolute error of phase-field for both increments. We observe a healthy reduction in the loss function, and we obtain a very good agreement between the ground truth and predicted phase-field across the entire domain. 

\section{Numerical examples}
\label{Section:Results}

In this section we demonstrate the results of our numerical implementation. Following the one-time training, we use the same PICNN to implement IFENN across several test cases. These include several variations of the single-notch and double-notch problem, and they are presented in detail below. 

\subsection{Single-Notch specimen under Tension (SNT)}
\label{Section:Results_SNT}

\begin{figure}[b!]
    \centering
    \includegraphics[width=0.9\textwidth]{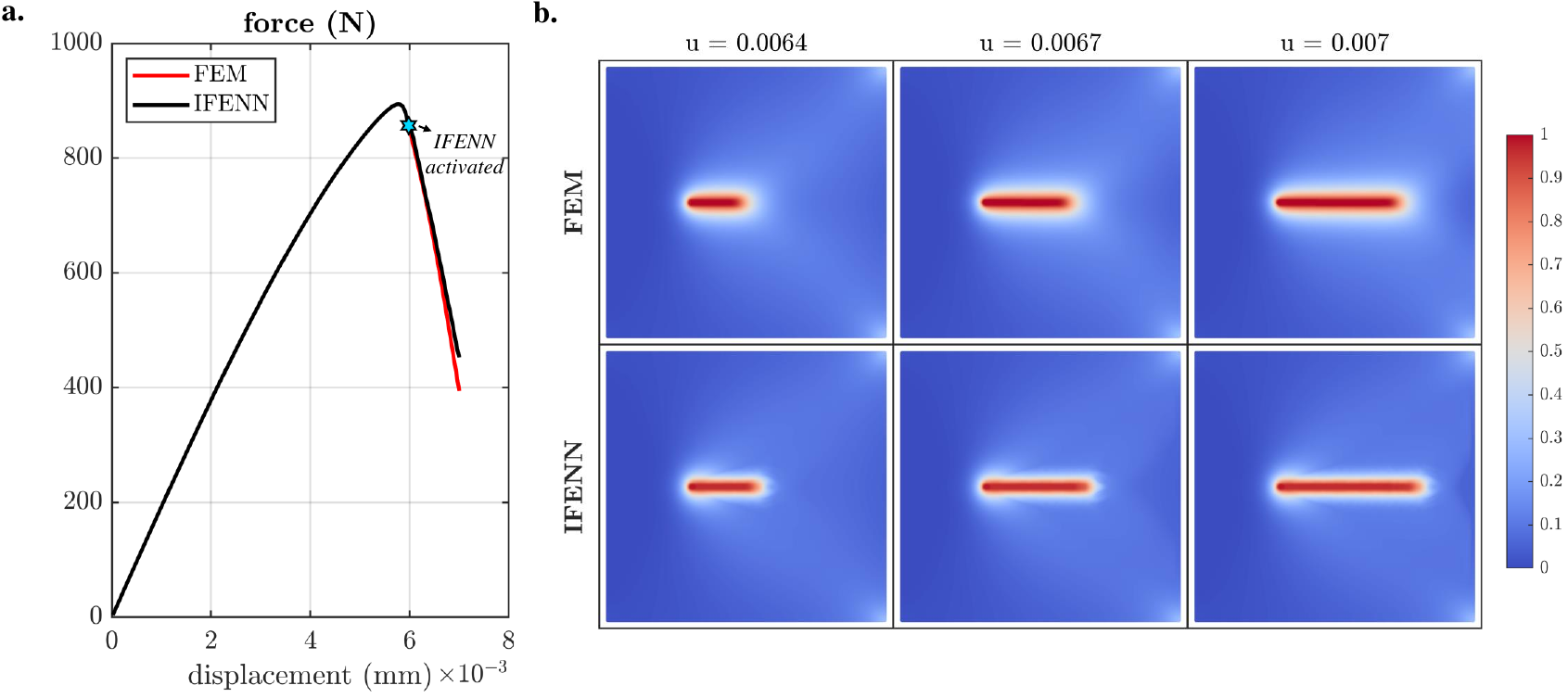}
    \caption{Comparison between FEM and IFENN for the SNT1 – I (350 increments) problem. {\bf{a.}} Vertical reaction force-displacement curves. {\bf{a.}} Phase-field contours from FEM (top) and IFENN (bottom) at various load levels. }
    \label{Figure_SNT_350inc_RFcontours}
\end{figure}

We begin our investigation on the SNT problem, the geometric and loading details of which were discussed in the previous section. The first SNT variation is the SNT1-I model, which was created previously for the PICNN training and we now simulate with both FEM and IFENN. We denote that IFENN is activated when the maximum nodal phase-field value reaches $\phi_{nodal,max} = 0.99$. Additionally, per our discussion in Section \ref{Section:IFENN_with_PICNNs}, the PICNN predictions are post-processed at every increment with a convolution-based Gaussian smoothing filter ($k = 5$, $\sigma = 2$). Fig. \ref{Figure_SNT_350inc_RFcontours}a illustrates the evolution of the vertical reaction force with respect to the imposed displacement for both FEM and IFENN. An excellent agreement between the two curves is observed. Additionally, Fig. \ref{Figure_SNT_350inc_RFcontours}b depicts phase-field contours at selected load increments. In all cases IFENN yields contours which are highly consistent with those from FEM, with a tendency to produce marginally thinner and more elongated cracks. 

\begin{figure}[H]
    \centering
    \includegraphics[width=1\textwidth]{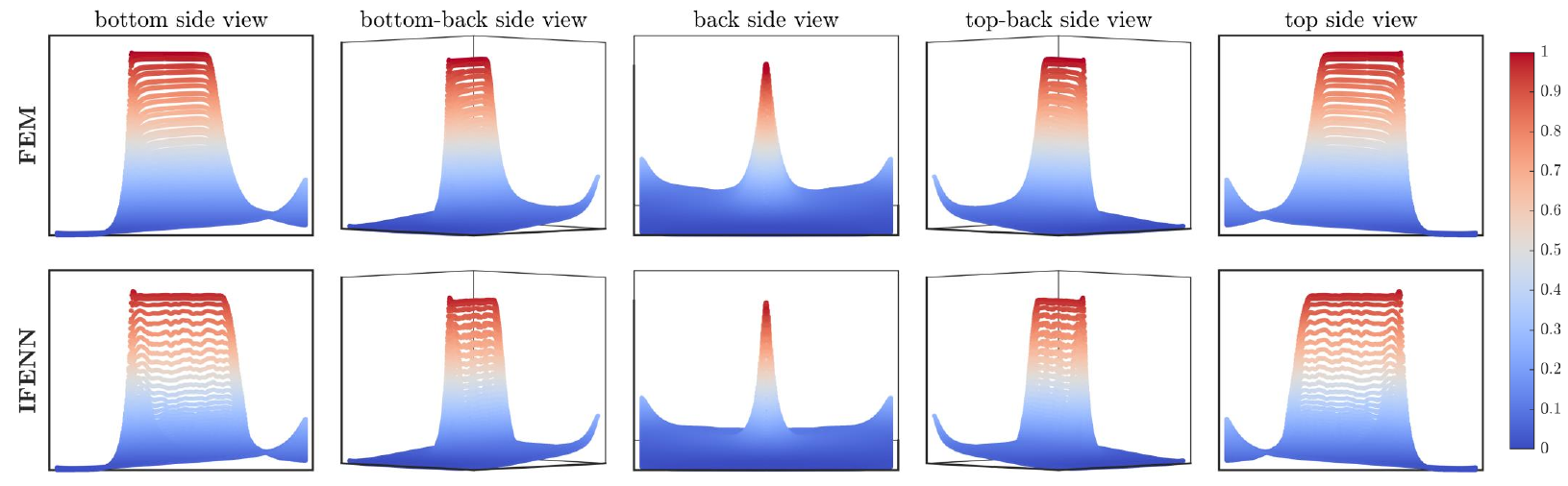}
    \caption{Comparison between FEM and IFENN for the SNT1 – I (350 increments): close-up 3D views of $\phi$ contours at inc = 340. The azimuth-elevation pairs for each angle are: bottom side (0$^{\circ}$,0$^{\circ}$), bottom back side (-45$^{\circ}$,3$^{\circ}$), back side (-90$^{\circ}$,10$^{\circ}$), top back (-135$^{\circ}$,3$^{\circ}$), top side (-180$^{\circ}$,0$^{\circ}$).}
    \label{Figure_SNT_350inc_contours_3D}
\end{figure}

The quality of the IFENN phase-field predictions can be more clearly appreciated when the results are visualized in 3D. Fig. \ref{Figure_SNT_350inc_contours_3D} presents 3D snapshots of the phase-field profile at the $340^{th}$ load increment from both FEM and IFENN, viewed from five different angles: bottom, bottom-back, back, top-back and top side. With both solvers we observe two clear and strong patterns: the $\phi$-profile is symmetric with respect to the crack centerline, and the Gauss-point values on any element row along the crack trajectory show a remarkable similarity. While both features are expected in a standard FEM analysis on a structured mesh, they are by no means guaranteed for a hybrid solver that relies on a neural network. In fact, the accuracy of NN-based predictions can be compromised by several parameters, and due to the strong path-dependence of phase-field propagation, even minor early-stage errors and non-symmetric predictions may result to accumulated error and unrealistic paths. In this case, the PICNN-based IFENN yields a smooth and symmetric crack opening which does not deviate from the expected trajectory. 

\begin{figure}[b!]
    \centering
    \includegraphics[width=0.9\textwidth]{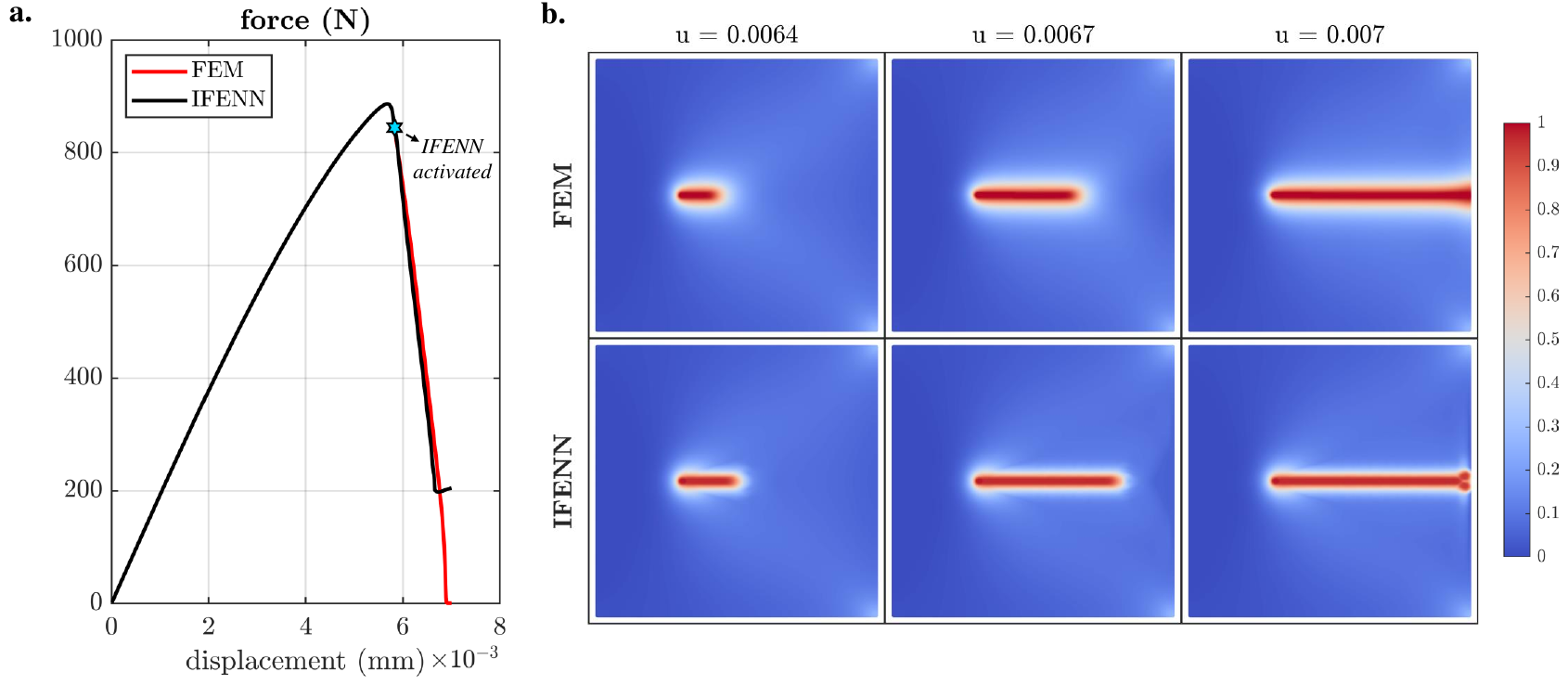}
    \caption{Comparison between FEM and IFENN for the SNT1 – II (700 increments) problem. {\bf{a.}} Vertical reaction force-displacement curves. {\bf{a.}} Phase-field contours from FEM (top) and IFENN (bottom) at various load levels. }
    \label{Figure_SNT_700inc_RFcontours}
\end{figure} 

\begin{figure}[H]
    \centering
    \includegraphics[width=0.9\textwidth]{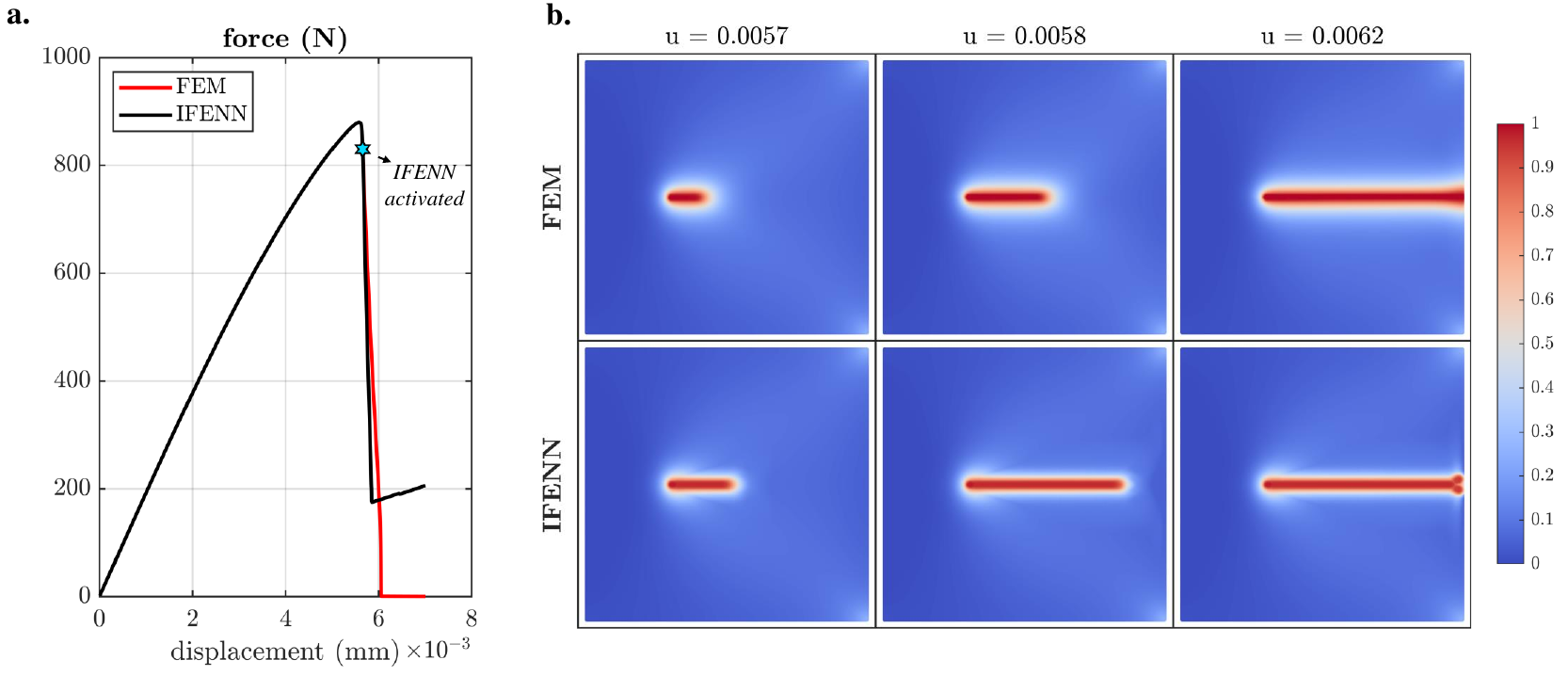}
    \caption{Comparison between FEM and IFENN for the SNT1 – III (1500 increments) problem. {\bf{a.}} Vertical reaction force-displacement curves. {\bf{a.}} Phase-field contours from FEM (top) and IFENN (bottom) at various load levels. }
    \label{Figure_SNT_1500inc_RFcontours}
\end{figure}

Next, we showcase the method's flexibility to use the same PICNN with different load stepping algorithms. To this end, we consider a second load case (\textit{II}), in which the analysis is carried out using 700 equidistant increments of $\Delta u = 1 \times 10^{-5}$ mm, as well as a third case (\textit{III}), where the load is first increased over 500 increments with $\Delta u = 1 \times 10^{-5}$ mm, followed by an additional 1000 increments with a finer step $\Delta u = 1 \times 10^{-6}$ mm. The resulting models are termed SNT1-II and SNT1-III, and we report the results in Fig. \ref{Figure_SNT_700inc_RFcontours} and Fig. \ref{Figure_SNT_1500inc_RFcontours} respectively. In both cases IFENN yields a reaction force that is almost identical to FEM, while the tendency to produce slightly thinner and more elongated cracks at the same load levels is still present. Here we note that the reaction forces do not fully drop to zero, and instead an artificial stiffening at the final part of these curves is observed. This is an artifact of the PICNN's $\phi$ predictions being post-processed with the Gaussian smoothing filter. This process diffuses the predicted $\phi$-profile and therefore slightly decreases its peak values from $\phi_{max} \approx$ 1.0 to $\phi_{max} \approx$ 0.95-0.96. This characteristic is consistent across all IFENN simulations and can be clearly seen in the bottom row of plots in Fig. \ref{Figure_SNT_350inc_contours_3D}: at the notch-tip $\phi_{max}$ is almost 1, while after IFENN is activated the remaining peak values are slightly lower. This numerical issue yields a residual stiffness even when the crack has fully propagated, hence the ascending branches in the last part of the reaction curves in Fig. \ref{Figure_SNT_700inc_RFcontours}a and \ref{Figure_SNT_700inc_RFcontours}b. This minor limitation is further discussed in the context of future enhancements in Section \ref{Section:Conclusions}.

\begin{figure}[b!]
    \centering
    \includegraphics[width=0.75\textwidth]{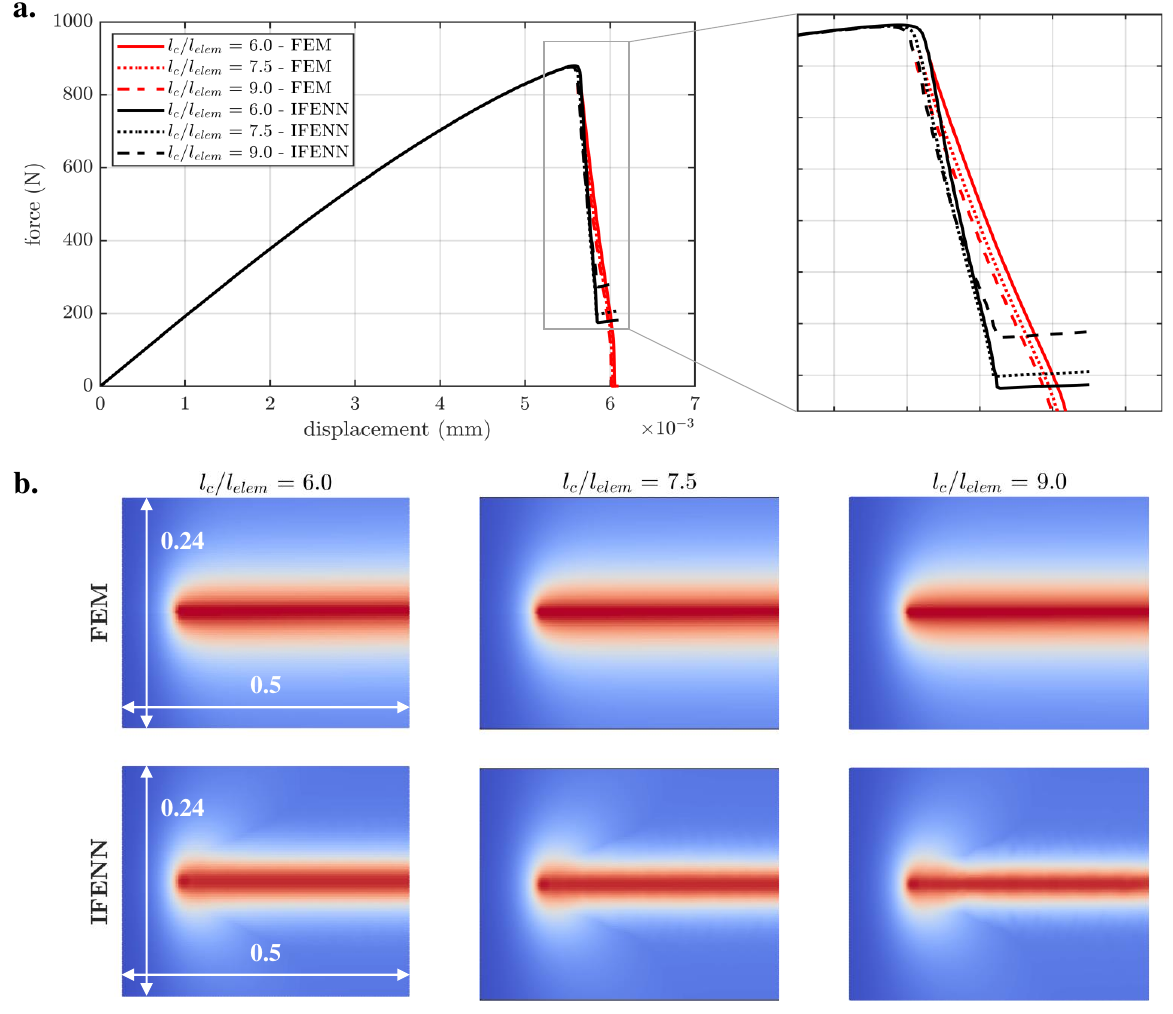}
    \caption{Comparison between FEM and IFENN for varying characteristic-to-element length ratios for the SNT problem. {\bf{a.}} Reaction force curves. {\bf{b.}} Phase-field contours in a zoomed-in region: {-0.3 $<$ x $<$ 0.2} $\&$ {-0.12 $<$ y $<$ 0.12}.}
    \label{Figure_SNT_lclelem_RFcontours}
\end{figure} 

\begin{figure}[b!]
    \centering
    \includegraphics[width=0.8\textwidth]{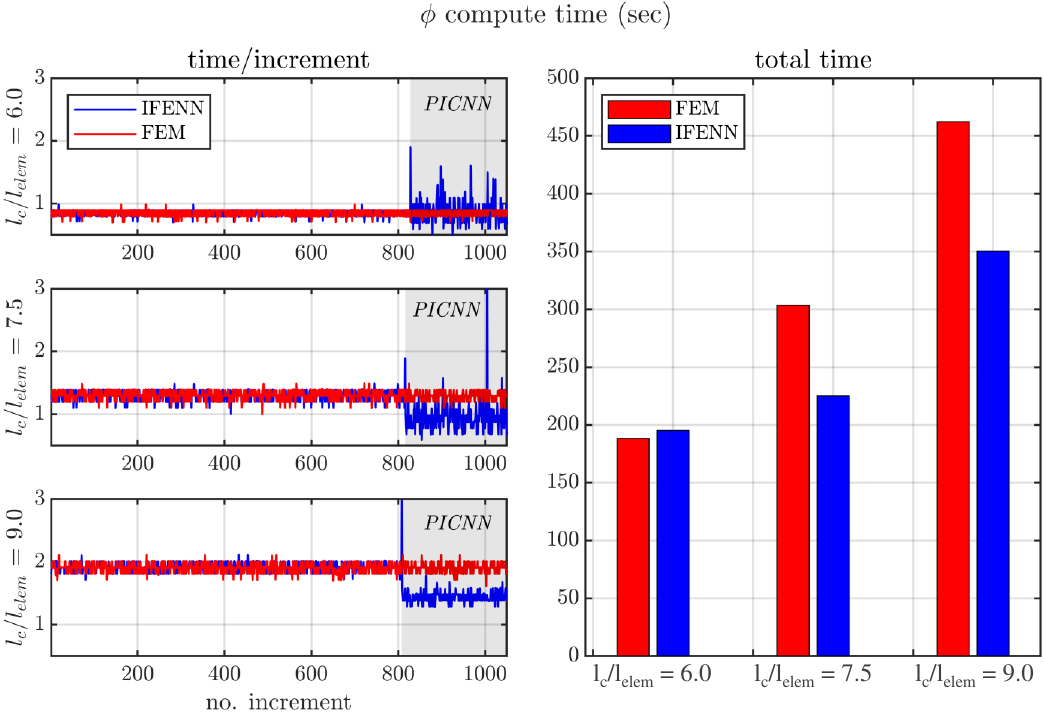}
    \caption{Computational time for the phase-field variable as the ratio of characteristic length to element size increases. The left plot shows time per increment, and the right plot presents the cumulative computation time during the propagation stage. IFENN demonstrates superior efficiency over FEM with increasing mesh resolution.}
    \label{Figure_SNT_lclelem_proptimes}
\end{figure} 

The last investigation of the SNT problem involves a study on progressively finer meshes, with the goal to demonstrate the framework's ability to utilize the same network across different mesh resolutions. The domain is further discretized using 61504 and 90000 square finite elements, and the resulting models are termed \textit{SNT2} and \textit{SNT3} with a characteristic-to-element length ratio of $l_{c} / l_{elem}$ = [7.5, 9.0] respectively. We perform the FEM and IFENN simulations on these models and report the results in Fig. \ref{Figure_SNT_lclelem_RFcontours}, capping the reaction force graph at u = 6.1 $\times 10^{-3}$ mm for clarity. First, we emphasize that IFENN is able to perform such simulations. This occurs despite the PICNN's pixel-based input in SNT2 and SNT3 analyses being 496 $\times$ 496 and 600 $\times$ 600 respectively, which is different than the 400 $\times$ 400 training input (SNT1). This feature is attributed to the PICNN design, which consists exclusively of convolution layers without flatten or dense layers, and proves its applicability to arbitrarily designed square domains. Second, we observe a very good agreement between IFENN and FEM across all models, as evidenced both by the reaction graph and the phase-field contours. Also, a close observation of the results shows evidence of an interesting trend: as $l_{c}$ remains constant and the mesh is more refined, FEM produces a constant crack width (as expected), while IFENN yields slightly thinner cracks. This feature can be seen by the contour plots of Fig. \ref{Figure_SNT_lclelem_RFcontours}b, and although it is only marginal, it points to an interesting direction: PICNN does not seem to follow the \textit{absolute value} of $l_{c}$. Instead, learning on a pixel-based spread of $\phi$ seems to dictate a PICNN predictive behavior that is based on the \textit{ratio} of the characteristic-to-element length. Essentially, the PICNN does not adhere to the physical length scale. Instead, it seems to learn during training the number of pixels that the diffused $\phi$ should spread over, and when used for simulations with smaller elements, this is translated to thinner predicted cracks. This hypothesis warrants a dedicated study and is considered beyond the scope of this article. 

Finally, we conclude this example by comparing the computational times for the phase-field variable between IFENN and FEM across the three models. The results of this comparison are presented in Fig. \ref{Figure_SNT_lclelem_proptimes}. The plots on the left correspond to the time-per-increment, while the graph on the right shows the cumulative computation time for $\phi$ in the propagation stage - the stage where the hybrid solver is activated. A clear trend of computational savings is observed for IFENN over FEM as the mesh resolution increases, highlighting the proposed framework's applicability and efficiency for large-scale, finer-mesh simulations.

\subsection{Double-Notch specimen under Tension (DNT)}
\label{Section:Results_DNT}

\begin{figure}[H]
    \centering
    \includegraphics[width=0.35\textwidth]{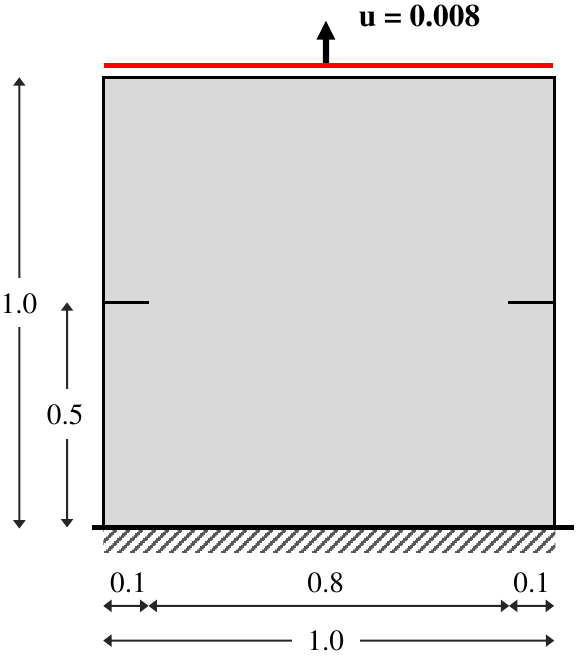}
    \caption{Geometric and loading/boundary condition details for the symmetric double-notch tension (SDNT) problem}
    \label{Figure_SDNT_schematic}
\end{figure}

In this example, we showcase that the same PICNN can be readily used to model the propagation of phase-field fracture from more than one front. To this end, we consider two different idealizations of a double-notch specimen subjected to tensile loading. The geometry and loading/boundary conditions of the first model are shown in Fig. \ref{Figure_SDNT_schematic}. It is a square domain with the same dimensions as the SNT model ($l_{x}$ = $l_{y}$ = 1.0 mm), fixed along the bottom edge and subjected to a uniform tensile displacement $u = 0.008$ along the top edge. Two cracks are introduced symmetrically on the left and right sides of the specimen at the same vertical position, forming the Symmetric Double-Notch Tension (SDNT) model. A square mesh of $1^{st}$-order quadrilateral finite elements is applied, with each element having a length of $l_{elem} = 0.005$ mm. The same material parameters as in the SNT problem are adopted. We also emphasize that the trained PICNN from Section \ref{Section:One_tine_training} is used without any additional modifications. 

\begin{figure}[t!]
    \centering
    \includegraphics[width=1\textwidth]{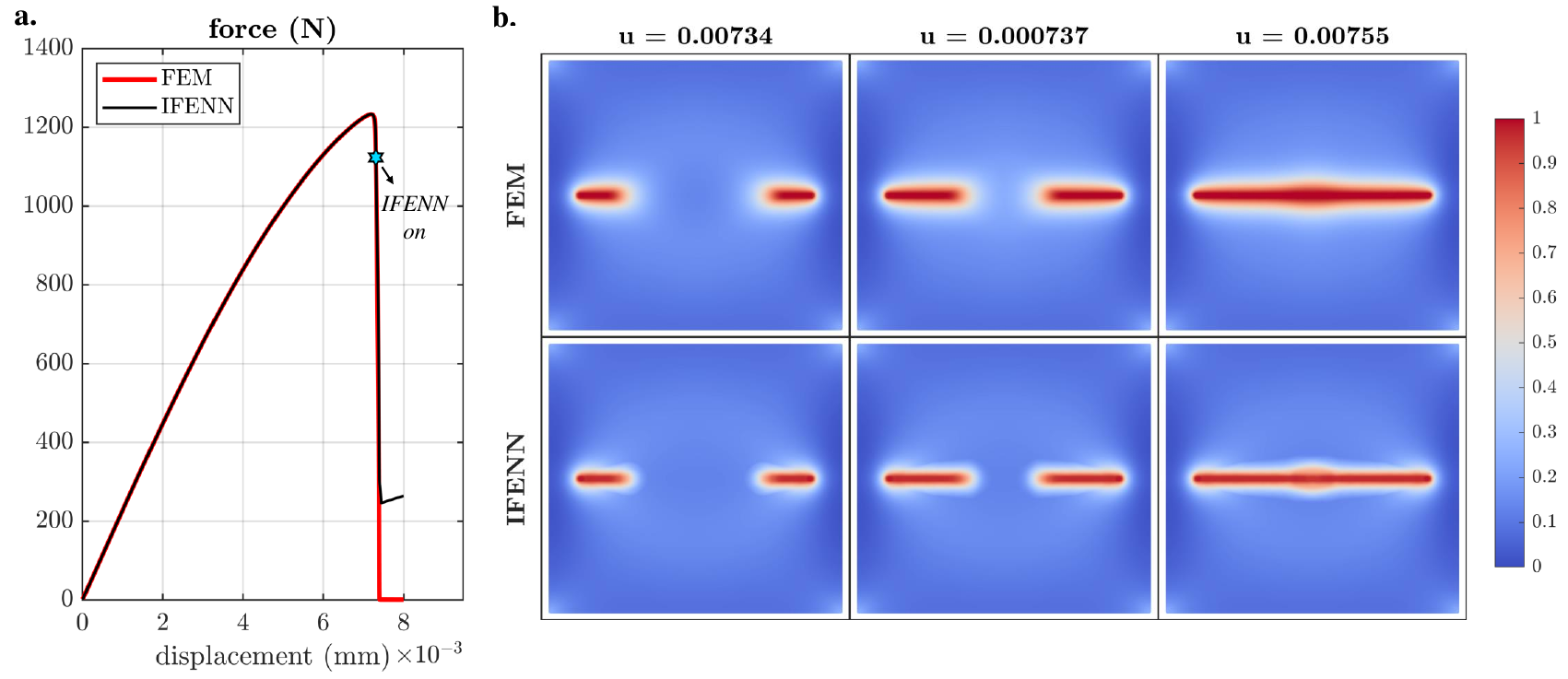}
    \caption{Comparison between FEM and IFENN for the SDNT problem. {\bf{a.}} Vertical reaction force-displacement curves. {\bf{b.}} Phase-field contours from FEM (top) and IFENN (bottom) at various load levels.}
    \label{Figure_DNT2_RFcontours}
\end{figure} 

The results of the FEM vs IFENN comparison for the SDNT problem are shown in Fig. \ref{Figure_DNT2_RFcontours}. First, we observe an almost exact coincidence of the vertical reaction forces following the activation of the hybrid solver. The ascending branch observed at end of the IFENN force curve once again arises from the previously discussed issue of phase-field not reaching the exact value of 1. More importantly, we observe in Fig. \ref{Figure_DNT2_RFcontours}b that IFENN predicts the formulation of both cracks and accurately tracks their trajectory until their full coalescence at the center. This is a remarkable finding, given that the PICNN was never trained on a second crack; yet, it infers its presence through the high H values in that region and translates this information into the corresponding phase-field. In fact, we note that both notches measure only 0.1 mm, whereas the notch used for training had a length of 0.3 mm. As a result, $both$ cracks propagate from different locations than the one seen in the training geometry. The fact that IFENN can predict the formulation of both cracks, offers the first strong evidence on the spatial generalization capability and geometric-invariance character of the underlying network.

\begin{figure}[b!]
    \centering
    \includegraphics[width=0.55\textwidth]{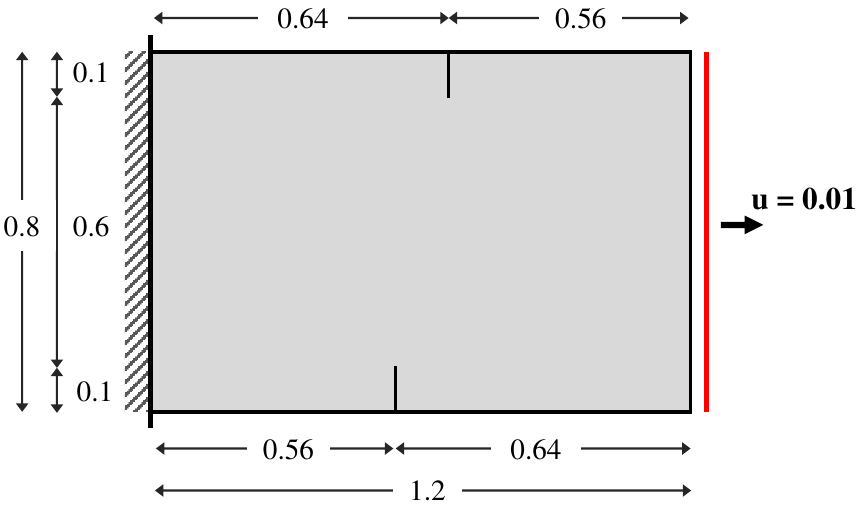}
    \caption{Geometric and loading/boundary condition details for the asymmetric double-notch tension (ADNT) problem}
    \label{Figure_ADNT_schematic}
\end{figure} 

In the following example, we tackle a significantly more challenging variation of the DNT problem by introducing three major modifications - each is designed to rigorously test the generalization capabilities of IFENN to unseen scenarios. First, we rotate both the notches and the loading direction, shifting from the previously studied horizontal propagation to a vertical orientation. This is in stark contrast with the previous examples and tests the ability of the PICNN to extrapolate to unseen directions. Second, we introduce an eccentric offset for both cracks with respect to their centerline. This setup is intended to evaluate the ability of IFENN to predict the propagation of asymmetric cracks and track their diagonal coalescence at the center. Third, we alter the domain dimensions from a square configuration ($l_{x} = l_{y} = 1$ mm) to a rectangular one ($l_{x} = 1.2$, $l_{y} = 0.8$ mm), with the goal to establish the generalizability of the PICNN to an arbitrarily designed - and unseen during training - rectangular domain. The material and mesh parameters are identical to the SDNT problem. This resulting model is termed Asymmetric Double Notch under Tension (ADNT) and a schematic is shown in Fig. \ref{Figure_ADNT_schematic}.

\begin{figure}[H]
    \centering
    \includegraphics[width=1\textwidth]{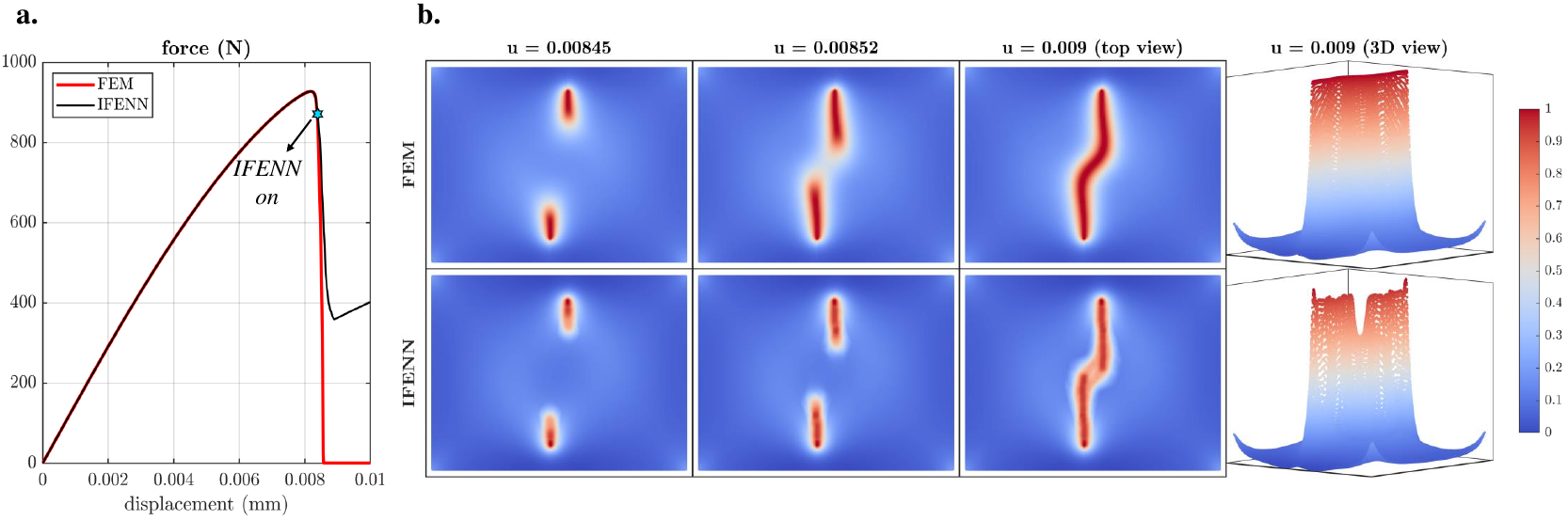}
    \caption{Comparison between FEM and IFENN for the ADNT problem. {\bf{a.}} Horizontal reaction force-displacement curves. {\bf{b.}} Phase-field contours from FEM (top) and IFENN (bottom) at various load levels.}
    \label{Figure_DNT3_RFcontours}
\end{figure} 

The results of the ADNT study are presented in Fig. \ref{Figure_DNT3_RFcontours}. The overall conclusions are consistent with those from the SDNT case, but we emphasize that here they are drawn in a substantially more difficult context. Overall, IFENN successfully predicts the formulation of both asymmetric vertical cracks and accurately follows their formulation until their diagonal merge at the center. It also captures the sharp drop of the horizontal reaction force. However, once the two cracks meet, the phase-field values at in the coalescence region stagnate around 0.7-0.75, as seen more clearly in the last column of plots in Fig. \ref{Figure_DNT3_RFcontours}b. This prevents the reaction force from fully dropping to zero. Nonetheless, the results in this figure provide the strongest yet evidence of the PICNN-IFENN's ability to generalize in unseen scenarios: the proposed scheme identifies and monitors the evolution of cracks which are completely different in terms of number, location and orientation than those seen training, while it remains applicable to arbitrarily rectangular domains.




\section{Conclusions}
\label{Section:Conclusions}

In this work, we develop a novel IFENN-based approach to model phase-field fracture propagation. The latter is an extremely challenging computational task, and our major contribution lies in carefully devising a novel network setup and associated hybrid solver that overcomes several fundamental and long-standing limitations in the literature of hybrid FEM-ML modeling. The main novelties and conclusions of this work are summarized below:

\begin{itemize}
    
    \item We adopt a purely unsupervised, 100$\%$ physics-based training approach for the NN-component of IFENN, which is a physics-informed CNN. This approach eliminates the need for specialized and heavy data-driven (labeled) training data.  

    \item By relying solely on the pixel-based representation of the $H$-$\phi$ mapping, we do not need to account for temporal features in both the training (offline) and inference (online) stage. This eliminates the need for fixed-length $Seq2Seq$ or temporal inference architectures, and introduces a new modeling paradigm in our ability to model path-dependent phenomena. 

    \item By using an ultra-low training dataset, which relies on the strain energy density profiles from only two load increments, we achieve exceptionally rapid training times ($\approx$ 5 minutes only). This shows that a super-fast training stage is both feasible and effective.
        
    \item The pixel-based $H-\phi$ mapping unlocks transformative modeling capabilities: we use the same trained PICNN across completely different model setups, including arbitrary rectangular domains, multiple interacting cracks, different mesh resolutions and varying load directions - all of which are unseen during training. In all cases, IFENN yields excellent agreement with the FEM solution, thus achieving excellent generalization performance.
    
\end{itemize}

The presented work builds on earlier IFENN research and offers a substantial advancement in the field of hybrid FEM-ML modeling, particularly for phase-field fracture simulations. It also opens the door for further research towards tangible improvements, aiming for more flexible, versatile, and efficient implementation. A major area for future enhancement involves the consideration of the damage nucleation and initiation stage into both the training and online stage, enabling the modeling of the entire fracture process. We are also actively expanding the method's applicability to irregular, non-rectangular geometries, which will drastically broaden the scope of this work. Another expansion could address the current reliance on uniformly-spaced structured meshes, as dictated by the pixel-based representation requirement of PICNNs. While our current approach offers a straightforward solution to the issue, it introduces an excessive number of elements in regions away from the fracture process zone and unnecessarily increases the cost. This can be addressed by employing a coarse unstructured mesh in those regions, combined with appropriate interpolation techniques to maintain the grid-like representation of the kinematic variables. Also, we observed the trained PICNN's tendency to preserve the characteristic-to-element length ratio rather than the characteristic length itself. This reflects a limitation that is akin to sequence models' lack of consideration of physical time, and indicates another interesting area of future improvement. Finally, the Gauss smoothing filter yields residual stiffness at the end of the simulation, and a potential remedy to this non-critical issue could be an appropriate scaling of the phase-field profile after the diffusion step.

\section*{Data availability}
\label{Section:Data_Availability}

All the code and data used in this work will be made publicly available upon publication of the article.

\newpage
\bibliography{bibliography}


\end{document}